\documentclass[final,5p,times,twocolumn]{elsarticle} 
\DeclareMathAlphabet{\mathpzc}{OT1}{pzc}{m}{it}
\usepackage{lineno}
\usepackage{setspace}
\usepackage{graphicx}
\usepackage{amssymb}
\usepackage{amsmath}
\usepackage{amsthm}
\usepackage[colorlinks=true]{hyperref}
\usepackage[all]{hypcap}
\usepackage{mathrsfs}

\usepackage[percent]{overpic}
\usepackage{fixltx2e}
\usepackage{epsfig}
\usepackage{verbatim}
\usepackage{multirow}
\usepackage{float}
\usepackage{array}

\usepackage{threeparttable}
\newcommand{\minitab}[2][l]{\begin{tabular}{#1}#2\end{tabular}}

\journal{Nuclear Instruments and Methods A}
\begin{document}
\begin{frontmatter}
\title{GEANT4 simulation of the neutron background of the C$_6$D$_6$ set-up for capture studies at n\_TOF}

\author[a]{P.~\v{Z}ugec\corref{cor1}}
\ead{pzugec@phy.hr}

\author[b]{N.~Colonna}
\author[a]{D.~Bosnar}
\author[c]{S.~Altstadt}
\author[d]{J.~Andrzejewski}
\author[e]{L.~Audouin}
\author[b]{M.~Barbagallo}
\author[f]{V.~B\'{e}cares}
\author[g]{F.~Be\v{c}v\'{a}\v{r}}
\author[h]{F.~Belloni}
% ...
\cortext[cor1]{Corresponding author. Tel.: +385 1 4605552}
\fntext[fn1]{www.cern.ch/ntof}

\begin{abstract}
The neutron sensitivity of the C$_6$D$_6$ detector setup used at n\_TOF for capture measurements has been studied by means of detailed GEANT4 simulations. A realistic software replica of the entire n\_TOF experimental hall, including the neutron beam line, sample, detector supports and the walls of the experimental area has been implemented in the simulations. The simulations have been analyzed in the same manner as experimental data, in particular by applying the Pulse Height Weighting Technique. The simulations have been validated against a measurement of the neutron background performed with a $^\mathrm{nat}$C sample, showing an excellent agreement above 1 keV. At lower energies, an additional component in the measured $^\mathrm{nat}$C yield has been discovered, which prevents the use of $^\mathrm{nat}$C data for neutron background estimates at neutron energies below a few hundred eV. The origin and time structure of the neutron background have been derived from the simulations. Examples of the neutron background for two different samples are demonstrating the important role of accurate simulations of the neutron background in capture cross section measurements.
\end{abstract}

\end{frontmatter}

\section{Introduction}
\label{sec:chap1}
Time-of-flight (TOF) measurements of neutron capture cross section rely on the detection of the prompt $\gamma$-rays emitted following the capture reaction. High-accuracy results require the minimization and precise determination of all possible sources of background. One of such components, the sample-related neutron background, originates from neutrons scattered by the sample, which are later captured or undergo inelastic reactions in the detectors or surrounding materials. The $\gamma$-rays produced in the process, if detected, give rise to a background event that can hardly be distinguished from a true capture reaction. Therefore, it is mandatory to minimize this background component by using detectors as insensitive to neutrons as possible. To this end, hydrogen-free detectors have been used for more than 40 years. C$_6$F$_6$ scintillators were originally employed by Macklin and Gibbons \cite{bibA} in measurements of neutron capture cross sections. However, this scintillator was later shown to still exhibit a non-negligible neutron sensitivity, resulting in rather large systematic uncertainties \cite{bibB}. The challenge was finally met by replacing C$_6$F$_6$ with C$_6$D$_6$ (deuterated benzene) liquid scintillator, which is still widely used in neutron facilities around the world.

At the neutron time of flight facility n\_TOF at CERN, a pair of specifically designed C$_6$D$_6$ detectors \cite{bibC} are being employed since a decade in measurements of radiative neutron capture cross sections of interest for both Nuclear Astrophysics and applications to emerging nuclear technologies \cite{nTOF1,nTOF2} (further details on the n\_TOF facility may be found in Ref. \cite{bibI}). To match the very low intrinsic neutron sensitivity of the scintillator liquid, care has been taken at n\_TOF in optimizing the materials and geometry of the various detector components. In particular, a large improvement has been achieved by using a carbon-fiber container, and Boron free window (quartz instead of the usual borosilicate). This has allowed to significantly reduce the neutron sensitivity of the detector, defined as the ratio $\varepsilon_n/\varepsilon_\gamma^\mathrm{max}$ of the neutron detection efficiency $\varepsilon_n$ and the maximal efficiency $\varepsilon_\gamma^\mathrm{max}$ for detecting a capture event, as indicated by Monte Carlo simulations and confirmed in dedicated measurements \cite{bibC}. A low value of the neutron sensitivity is fundamental for measuring isotopes characterized by very large scattering-to-capture cross section ratios, such as the stable magnesium isotopes \cite{bibQ}.

In a typical neutron capture cross section measurement the sample-related neutron background is generated not only by neutrons directly scattered towards the C$_6$D$_6$ detectors and captured therein, but also by neutrons scattered at the sample and then captured in other materials inside the experimental hall, such as the various components of the neutron beam line, the sample holder and detector supports, the floor, ceiling and lateral walls of the experimental area, etc., with the largest fraction of the background presumably generated by the material in the immediate vicinity of the irradiated sample.

The background related to the neutron sensitivity of the whole experimental setup, including the hall itself, is in principle easily accessible experimentally, by measuring the yield of a carbon sample. In fact, since $^\mathrm{nat}$C is characterized by a negligible capture cross section, its measured yield is expected to purely represent the background related to the neutron sensitivity of the capture setup. For this reason, a $^\mathrm{nat}$C sample is regularly measured in every capture experiment at n\_TOF, where its yield, suitably scaled, is used to determine and subtract the residual sample-related neutron background. However, while this procedure allows one to determine the average level of the neutron background, it does not provide information on its time structure. This is a problem in particular when measuring narrow resonances in the capture cross section. Furthermore,  there are doubts on the reliability of the measured $^\mathrm{nat}$C yield around thermal energy, due to other components, which are not related to the neutron sensitivity of the setup.

A complementary approach to the determination of the neutron background relies on the use of Monte Carlo simulations. Apart from  implementing a realistic software replica of the complex experimental apparatus -- all detector components, the neutron beam line, the walls of the experimental hall, etc. -- the main difficulty is related to the availability and the selection of appropriate physical models, in particular those for handling the hadronic interactions. Another problem is related to the very long computing time required for neutron transport, especially if one wants to determine the background as a function of neutron energy. At n\_TOF this is particularly cumbersome, considering that the neutron energy spans over twelve orders of magnitude, from thermal to the GeV neutron energy region. For all these reasons, only partial attempts have been performed in the past on simulating the neutron sensitivity of the whole capture setup. Recently, advances in the neutron physics part of GEANT4 simulation toolkit \cite{bibE,bibK}, combined with the capability of modern high-power computers, have offered the opportunity to perform an accurate and complete simulation of the neutron background in the whole neutron energy range.

We present in this paper the results of the simulations of the neutron background and its time dependence, performed for the first time for the complete experimental capture setup of n\_TOF. The results of the simulations have been validated in a specific energy range by comparison against experimental data obtained with a $^\mathrm{nat}$C sample. A careful analysis of the simulations has revealed an unexpected component, that questions the use of this sample for estimating the neutron background of other measurements at low energy. On the contrary, this component can actually provide important information on the integral cross section of the $(n,p)$ reaction on $^\mathrm{12}$C.

\begin{table*}[t!]
\caption{Overview of the physics list used for neutron transport with GEANT4 in this work.}
\label{tabx}
\centering
\begin{threeparttable}
\begin{tabular}{clll}
\hline\hline
\multirow{2}{*}{\textbf{Process}}&\multicolumn{2}{c}{\textbf{Model}}&\multicolumn{1}{c}{\multirow{2}{*}{\textbf{Cross section data}}}\\
\cline{2-3}&\minitab[c]{\textbf{Name}}&\minitab[c]{\textbf{Range}}&\\
\hline
\minitab[c]{Elastic\\scattering}&\minitab[l]{G4NeutronHPThermalScattering\\G4NeutronHPElastic\\G4HadronElastic}&\minitab[l]{$<$4 eV\\4 eV -- 20 MeV\\$>$20 MeV}&\minitab[l]{G4NeutronHPThermalScatteringData\tnote{d}\\G4NeutronHPElasticData\tnote{e}\\G4NeutronHPJENDLHEElasticData\tnote{f}\\G4BGGNucleonElasticXS\tnote{gh}}\\
\hline
\minitab[c]{Inelastic\\scattering}&\minitab[l]{G4BinaryCascade\tnote{a}\\G4CascadeInterface\tnote{a}\\G4TheoFSGenerator\tnote{b}}&\minitab[l]{$<$30 MeV\\30 MeV -- 10 GeV\\$>$10 GeV}&\minitab[l]{G4NeutronHPThermalScatteringData\tnote{d}\\G4NeutronHPInelasticData\tnote{e}\\G4NeutronHPJENDLHEInelasticData\tnote{f}\\G4BGGNucleonInelasticXS\tnote{gi}}\\
\hline
\minitab[c]{Capture}&\minitab[l]{G4NeutronHPCapture\\G4NeutronRadCapture}&\minitab[l]{$<$20 MeV\\$>$20 MeV}&\minitab[l]{G4NeutronHPCaptureData\tnote{e}\\G4HadronCaptureDataSet\tnote{g}}\\
\hline
\minitab[c]{Fission}&\minitab[l]{G4ParaFissionModel\tnote{c}\\G4LFission}&\minitab[l]{$<$60 MeV\\$>$60 MeV}&\minitab[l]{G4NeutronHPFissionData\tnote{e}\\G4HadronFissionDataSet\tnote{g}}\\
\hline\hline
\end{tabular}
\begin{tablenotes}
\item[a]Substitute: G4NeutronHPInelastic ($<$20 MeV) and G4NeutronHPThermalScattering ($<$4 eV)
\item[b] Using: G4StringChipsParticleLevelInterface, G4QGSMFragmentation and G4QGSModel\\ Substitute: G4GeneratorPrecompoundInterface, G4LundStringFragmentation and G4FTFModel
\item[c] Substitute: G4NeutronHPFission ($<$20 MeV)
\item[d] Declared range: $<$ 4 eV
\item[e] Declared range: $<$ 20 MeV
\item[f] Declared range: $\lesssim$3 GeV
\item[g] Declared range: $<$ 100 TeV
\item[h] Substitute: G4HadronElasticDataSet
\item[i] Closest substitute: G4NeutronInelasticCrossSection\\ Additional substitute: G4HadronInelasticDataSet
\end{tablenotes}
\end{threeparttable}
\end{table*}

The paper is organized as follows. The simulations are described in Section \ref{sec:chap2}, Section \ref{sec:chap3} deals with the analysis of the output data, and Section ~\ref{sec:chap4} presents the simulated neutron sensitivity of the whole experimental setup based on C$_6$D$_6$ detectors commonly used for neutron capture cross section measurements at n\_TOF. The energy dependence of the neutron sensitivity is analyzed, together with the origin of various components of the sensitivity. In Section \ref{sec:chap5}, the Monte Carlo results are validated by comparing the results of the simulations with experimental data for a $^\mathrm{nat}$C sample. Finally, examples of the effect of the simulated neutron background for two different samples are discussed.

\section{GEANT4 simulations}
\label{sec:chap2}

The simulations are based on GEANT4 version 9.6.p01. To set up the physics framework for the simulation, the necessary particles and physical processes had to be constructed. The list of general physical processes, like particle and radioactive decay, hadron interactions, etc., for all particles other than neutrons was set up from prearranged, standard packages. Special attention was paid to the neutron physics. As a first step, all processes assigned to neutron transport by any of the previous constructors were decoupled from the list. All required packages for handling the neutron physics were then specifically selected one by one, according to accuracy and availability within a given energy range. The selected neutron processes include elastic and inelastic scattering, neutron capture and neutron-induced fission. The physical models selected for a given process are listed in Table \ref{tabx}, together with the selected cross section datasets. The energy range for all models was set in accordance with the declared or the recommended range. The determination of the energy range of cross sections in a database was left to internal GEANT4 procedures, since not all of them contain data for all known nuclides.

GEANT4 offers a multitude of models for handling a physical process within a given energy range. Therefore, the choices made for this work are listed in Table \ref{tabx} explicitly. For all processes the G4NeutronHP package offers separate high-precision models, based on a direct sampling of the detailed cross section data. In general, for neutron-induced reactions up to 20 MeV, cross sections from the evaluated ENDF/B-VII library are used \cite{bibL,bibJ}.

In this context, inelastic and ($n$,charged particle) reactions represent special cases. Based on the comparison with the experimental data, in particular for the carbon sample, a combination of G4BinaryCascade and G4CascadeInterface (the so-called Bertini cascade) was selected within the entire energy range below 10 GeV. For fission processes, G4ParaFissionModel was preferred over G4HeutronHPFission because the latter does not produce fission fragments. However, it should be kept in mind that -- according to the online documentation \cite{bibY} -- G4ParaFissionModel does not produce as much of the delayed neutrons as G4NeutronHPFission model, which is not of a concern in the present work.

In Table \ref{tabx} the cross section datasets used in different energy ranges are also listed. The Barashenkov-Glauber-Gribov parameterization of inelastic cross sections has been preferred over that of Laidlaw-Wellisch, as suggested in Ref. \cite{bibZ}.

\begin{figure}[b!]
\includegraphics[width=1.\linewidth,keepaspectratio]{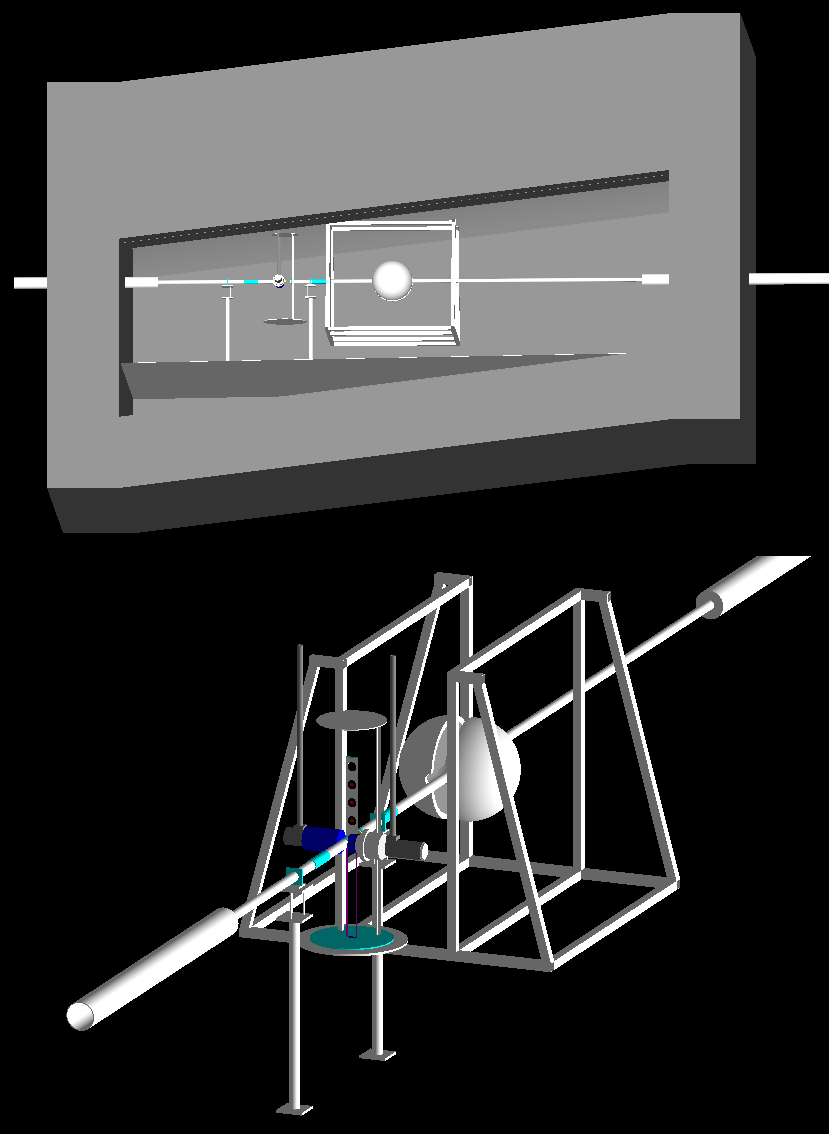}
\caption{Visualization of the geometric configuration used in the GEANT4 simulation. The upper segment shows the whole setup, surrounded by the concrete walls of the experimental area. The lower segment is a zoom of the detectors.}
\label{fig3}
\end{figure}

The geometry of the n\_TOF experimental hall, as implemented within the simulation code, is displayed in Fig. \ref{fig3}. The configuration exactly reproduces the experimental setup used in a recent measurement of the $^{58}$Ni neutron capture cross section \cite{bibD}. Two different detectors were used in that case (as commonly done in similar measurements): one specifically built at Forschungszentrum Karlsruhe (see \cite{bibC}), with a carbon-fiber cell for optimized neutron sensitivity, while the other one is a modified version of the detector available from Bicron. The detectors were mounted face-to-face, perpendicular to the beam direction, and slightly backwards relative to the sample. A carbon fiber sample exchanger was used for mounting the samples.

Care has been taken in accurately reproducing the layout of the hall, which is slightly tilted relative to the horizontal plane. The concrete walls of the experimental area have been included for a thickness of 1.5 meters. For the experiments involving intense neutron beams, it is important to consider the concrete walls because their hydrogen content is responsible for neutron moderation and subsequent capture with the emission of 2.2 MeV $\gamma$-rays. A false floor, made of thin aluminum plates mounted on a support grid, has also been implemented in the simulation, together with the Al neutron beam line, the carbon fiber sample holder and detector supports. Finally, as shown in Fig.~\ref{fig3}, the geometry includes a 4$\pi$ calorimeter made of BaF$_{2}$ crystals and the corresponding support structure made of aluminum, which is permanently installed in the experimental area. Furthermore, a detailed software replica of the detectors has been implemented in the simulations. The geometry was adopted from the simulations \cite{bibH} used for the weighting function, since this ensures that the Pulse Height Weighting Technique (PHWT) is consistently applied when analysing the simulations here discussed.

\begin{figure}[b!]
\includegraphics[angle=90,width=1.\linewidth,keepaspectratio]{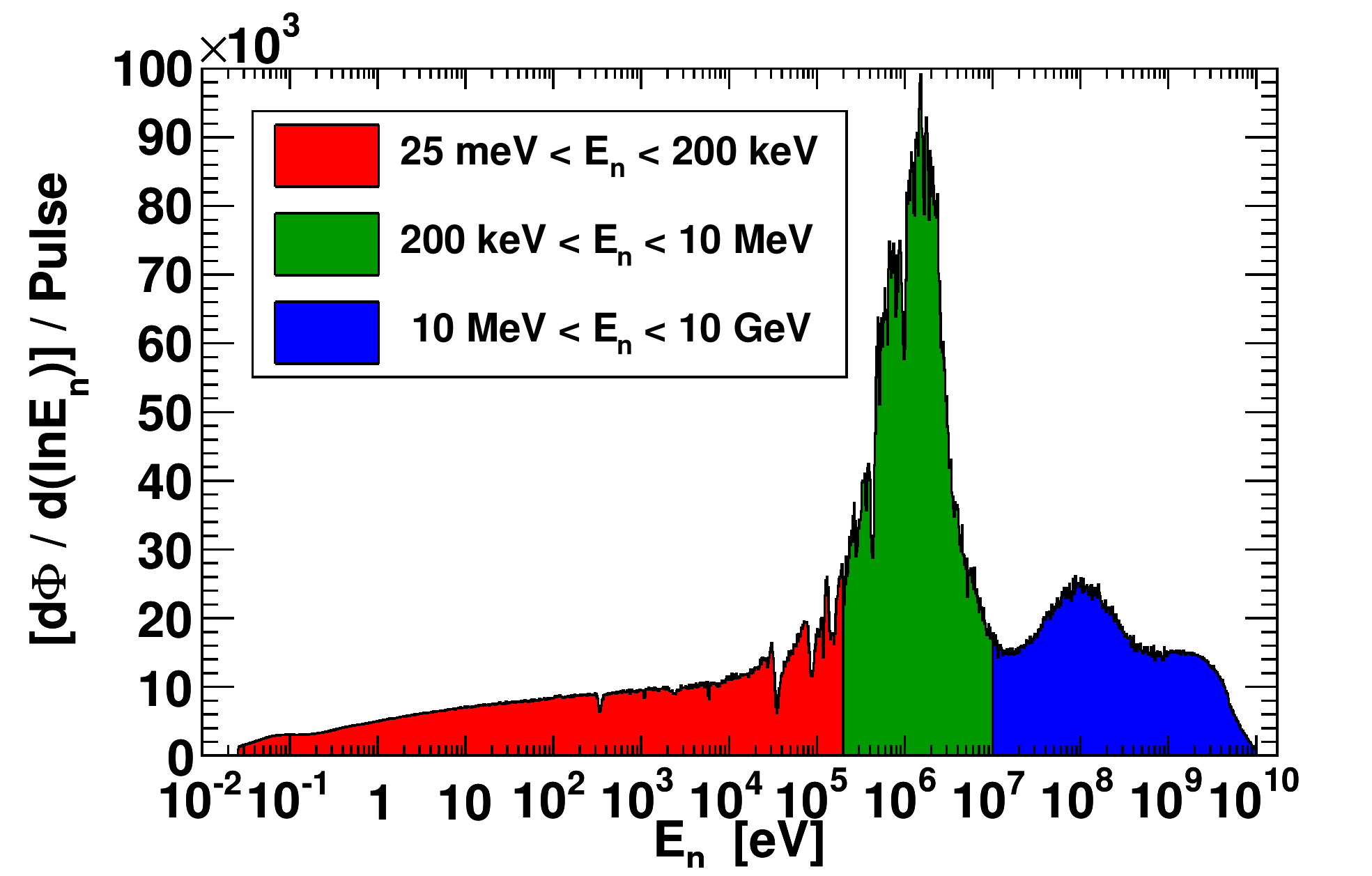}
\caption{(Color online) Energy dependent neutron flux from 2011, adopted for sampling the neutron energy in the present GEANT4 simulation. The differently colored energy intervals were handled by separate simulation runs, in order to optimize the CPU processing time.}
\label{fig4}
\end{figure}

In the simulations, the neutron energy is sampled from the spectrum of the n\_TOF neutron beam (see Refs. \cite{bibI,bibM} for details on the neutron flux). For samples with a diameter smaller than the neutron beam, the energy dependence of the beam interception factor (BIF) must be taken into account, in order to correctly determine the neutron flux incident on the sample. Figure \ref{fig4} shows the flux incident on the samples of 2 cm diameter, as used in the simulations. Since the measurements are available up to a neutron energy of 1 GeV, the results from the dedicated FLUKA simulations \cite{bibI} were used to extend the neutron flux up to 10 GeV. For an optimal management of the processing time, the neutron energy spectrum was divided in three regions. Separate simulation runs were performed in parallel for primary neutron energies from 25 meV to 200 keV, 200 keV to 10 MeV, and 10 MeV to 10 GeV, as indicated in Fig. \ref{fig4}. The relative portions of the total flux within these intervals , when the correction for the BIF dependence is applied, are 29.9\%, 45.0\% and 25.1\%, respectively. The neutron beam profile was assumed to be Gaussian, with experimentally determined horizontal and vertical widths of $\sigma_x=\sigma_y=7.5$~mm.

Since the used version 9.6.p01 of GEANT4 is not parallel-processing oriented by default, simulations were run in parallel on several PCs, even for the same part of the neutron spectrum, in order to accumulate as much statistics as possible. In this case, since the default random number generators in GEANT4 produce the same sequence of numbers, the starting seed in every run was randomized on the basis of a local starting time, so as to obtain a unique stream of output data for each run.

The aim of the simulation is to estimate the background caused by neutrons scattered off the sample. For this reason, only neutrons that undergo a non-negligible interaction with the sample are followed in the simulations (the background generated by the interaction of the neutron beam before or after the sample does not need to be simulated, since it is normally measured in runs without the samples). In the case of elastic scattering the condition for considering the interaction as non-negligible had to be set manually, requesting that the cumulative scattering angle of the neutron at the exit surface of the sample be greater than 0.2$^\circ$. Once a neutron is scattered by the sample, it is followed until it either decays, undergoes another interaction or reaches the boundary of the volume (called "mother volume") containing all other volumes. In case of an interaction or decay, all secondary particles generated are transported, until they are stopped or exit the world-volume. A default range cut of 1 mm was used for the production of secondary particles.

The output of the simulations consist of a large amount of detailed information on the mechanism and position of each particle interaction, as well as on the kinematics of the interaction. Considering the goal of the simulations, the main information regards the energy deposited in the C$_6$D$_6$ detectors, the type and energy of the particle that deposited it, and the time interval between the neutron generation and the earliest energy deposition inside the given C$_6$D$_6$ volume. Together with the original neutron energy, these quantities are sufficient to reconstruct the neutron background and its time structure, by applying the analysis procedure described later on. Other important information obtained from the simulations include the source of the background and the volume where it was generated.

\section{Analysis of the simulations}
\label{sec:chap3}

For consistency, the output data of the GEANT4 simulations are analyzed in the same way as the data collected during the measurements at n\_TOF. The first step is to assign to a given energy deposition in the C$_6$D$_6$ detector the corresponding time-of-flight. To this end, since neutrons in the simulations are generated at the entrance of the experimental area rather than at the spallation target, the time information provided by the simulation is added to the time-of-flight calculated from the original neutron energy, assuming the known distance between the spallation target and the experimental area (182 m). This total time-of-flight allows one to later assign the background event to a reconstructed neutron energy, in the same way as in the real measurement. In the analysis of the simulations the time-of-flight was limited to 96~ms, consistent with the actual width of the acquisition window used for the flash-ADC system at n\_TOF. However, for a realistic estimate of events with long delay (mostly associated with radioactive decays of activation products), the pulsed time-structure of the n\_TOF neutron beam was also reproduced in the code. To this end, the occurrence of a neutron pulse at a later time was simulated by using a Poisson distribution of bunches with a maximum repetition rate of 0.8 Hz (related to the  characteristics of the Proton Syncrotron accelerator) and an average repetition rate of 0.4 Hz (due to a limitation in the maximum power deliverable on the spallation target). The maximal time-of-flight  in the analysis of the simulations was set to 2 minutes, due to finite precision considerations. If, within this time, an event falls in a late neutron pulse, the time-of-flight is recalculated relative to the start of this new pulse, simulating a "wrap-around" process.

The second step regards the weighting of the deposited energy by means of the Pulse Height Weighting Technique commonly used in the analysis of capture data with C$_6$D$_6$ detectors. This technique, based on the original idea by Maier-Leibnitz \cite{bibA}, is applied in the analysis of the capture data in order to make the detection efficiency of these detectors independent of the de-excitation cascade path, i.e. of the energy and multiplicity of $\gamma$-rays emitted in the reaction. An in-depth description of the weighting function procedure adopted at n\_TOF may be found in Ref. \cite{bibH}. 

As previously mentioned, the weighting functions were determined for the same detector geometry as used in the present GEANT4 simulations. Adopting exactly the same geometrical setup is fundamental for a perfectly consistent comparison between simulated and experimental data. Thus the same weighting functions used for the experimental data have been applied to the simulated energy deposited in the C$_6$D$_6$ detectors, after convolution with the energy resolution experimentally determined with calibration $\gamma$-ray sources for each of the two detectors.

As a preliminary check of the reliability of the Physics list chosen for neutron transport in the present GEANT4 simulations, a control case was investigated, described in Ref. \cite{bibF}, showing excellent agreement with MCNP-4B and GEANT3 results.

\subsection{The $\gamma$-ray cascades}

\begin{figure}[b!]
\includegraphics[angle=90,width=1.\linewidth,keepaspectratio]{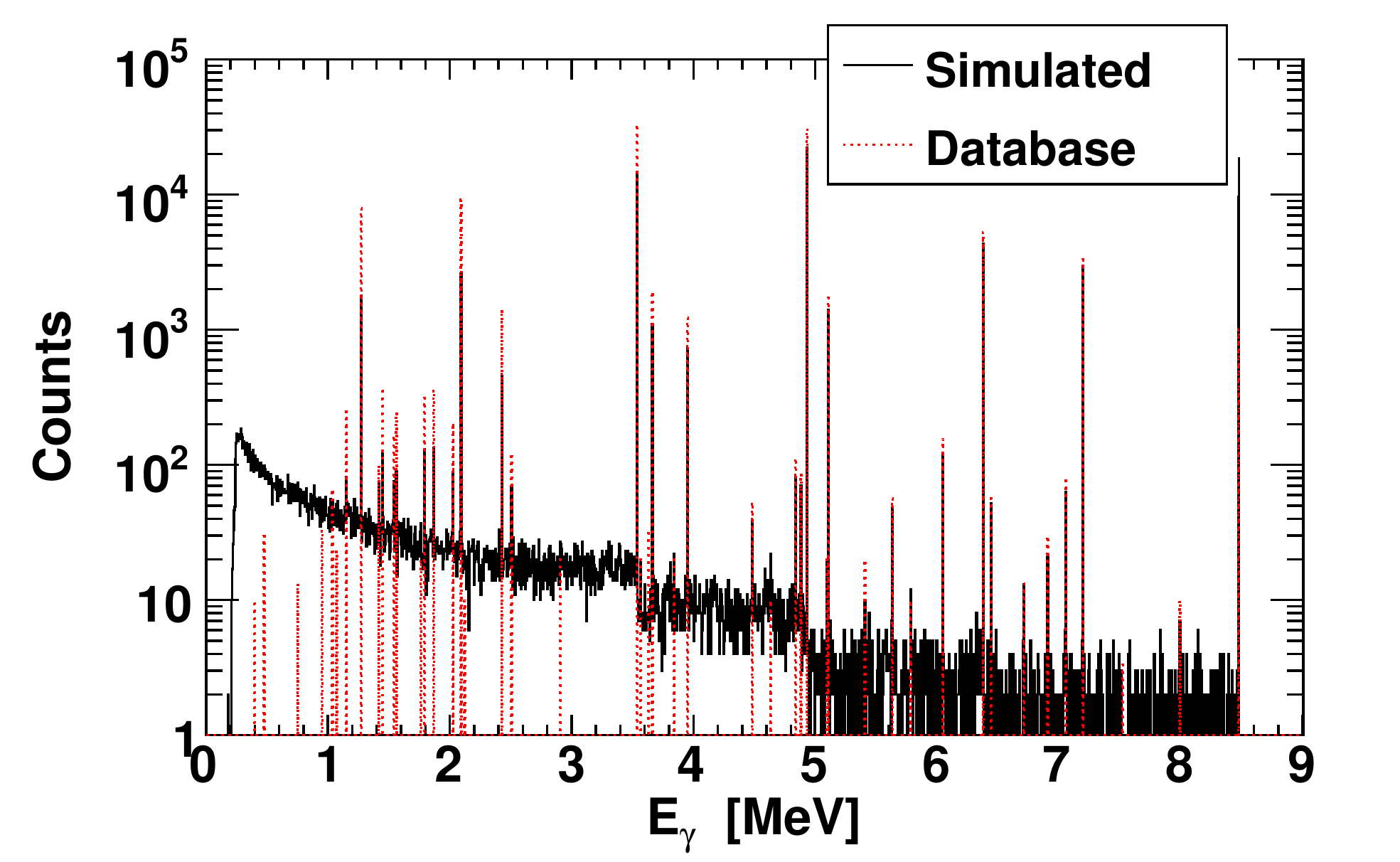}
\caption{(Color online) Simulated $\gamma$-ray spectrum from the cascade following thermal neutron capture on $^{28}$Si, compared to the tabulated $\gamma$-ray energies and intensities from Ref. \cite{bibN}.}
\label{fig5}
\end{figure}

The reliability of the $\gamma$-ray cascades generated within GEANT4 in a neutron capture reaction is an important issue that needs to be addressed. The main question is whether the generated cascades are realistic, i.e. if they are generated according to tabulated data (instead of being sampled from an artificial continuous distribution) and if the correlations between $\gamma$-rays are correctly taken into account. A detailed investigation was performed in order to address this issue. Dedicated GEANT4 simulations were performed for the neutron irradiation of a given isotope. Several isotopes, e.g. $^{23}$Na, $^{28}$Si, $^{56}$Fe, $^{58}$Ni, were selected for the present neutron sensitivity study. The energies of the $\gamma$-rays generated in the simulations of a capture reaction were recorded immediately upon their production, before they could interact with the surrounding material. Their spectra were then compared with the tabulated thermal neutron capture data from Ref. \cite{bibN}. $^{28}$Si has been selected as the clearest example due to the relatively low number of known capture $\gamma$-rays, which makes it easy to visually compare the spectra shown in Fig. \ref{fig5}. A close look at the narrow peaks in the simulated spectrum clearly reveals that $\gamma$-ray cascades are indeed generated according to the available tabulated data (if the tabulated data are not available for a given isotope, the continuum distribution is used \cite{physics}). However, the fact that $\gamma$-ray intensities are not accurately reproduced, together with the existence of the continuous spectrum outside the sharp peaks, is the unmistakable signature of the absence of correlations. It is implied that some  $\gamma$-rays are generated independently of the actual cascade path allowed by the nuclear structure. In fact, the energy of the final $\gamma$-ray emitted in a cascade is constrained by energy conservation, so that the sum of all $\gamma$-ray energies reproduces the excitation energy of the compound nucleus. Though this energy conservation requirement may be deactivated by user, it is essential for the Pulse Height Weighting Technique, in which the detection efficiency of a capture reaction is modified so to be independent on the cascade path and always equal to the excitation energy of the compound nucleus.

\begin{figure}[b!]
\includegraphics[angle=90,width=1.\linewidth,keepaspectratio]{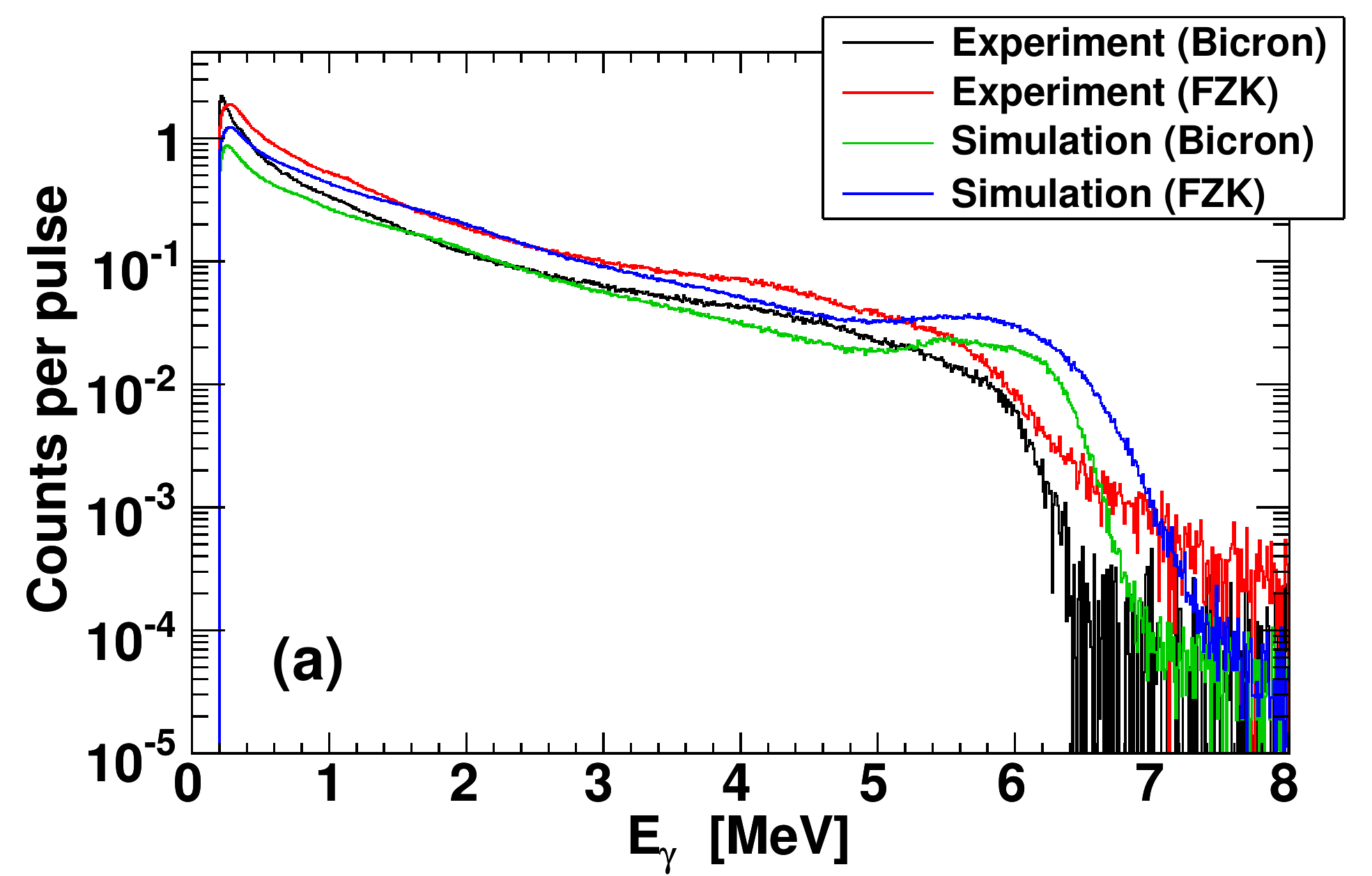}\\
\includegraphics[angle=90,width=1.\linewidth,keepaspectratio]{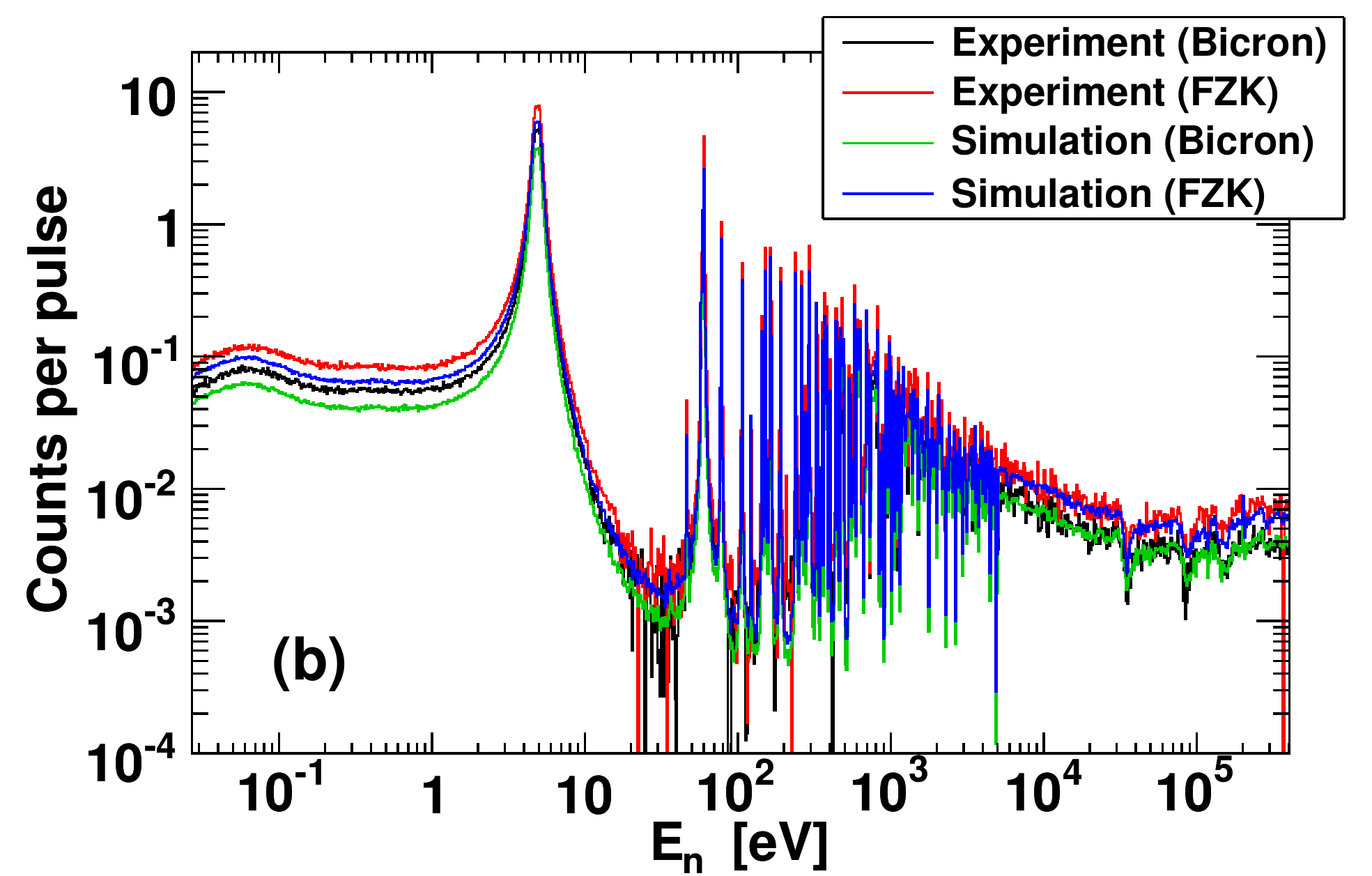}\\
\includegraphics[angle=90,width=1.\linewidth,keepaspectratio]{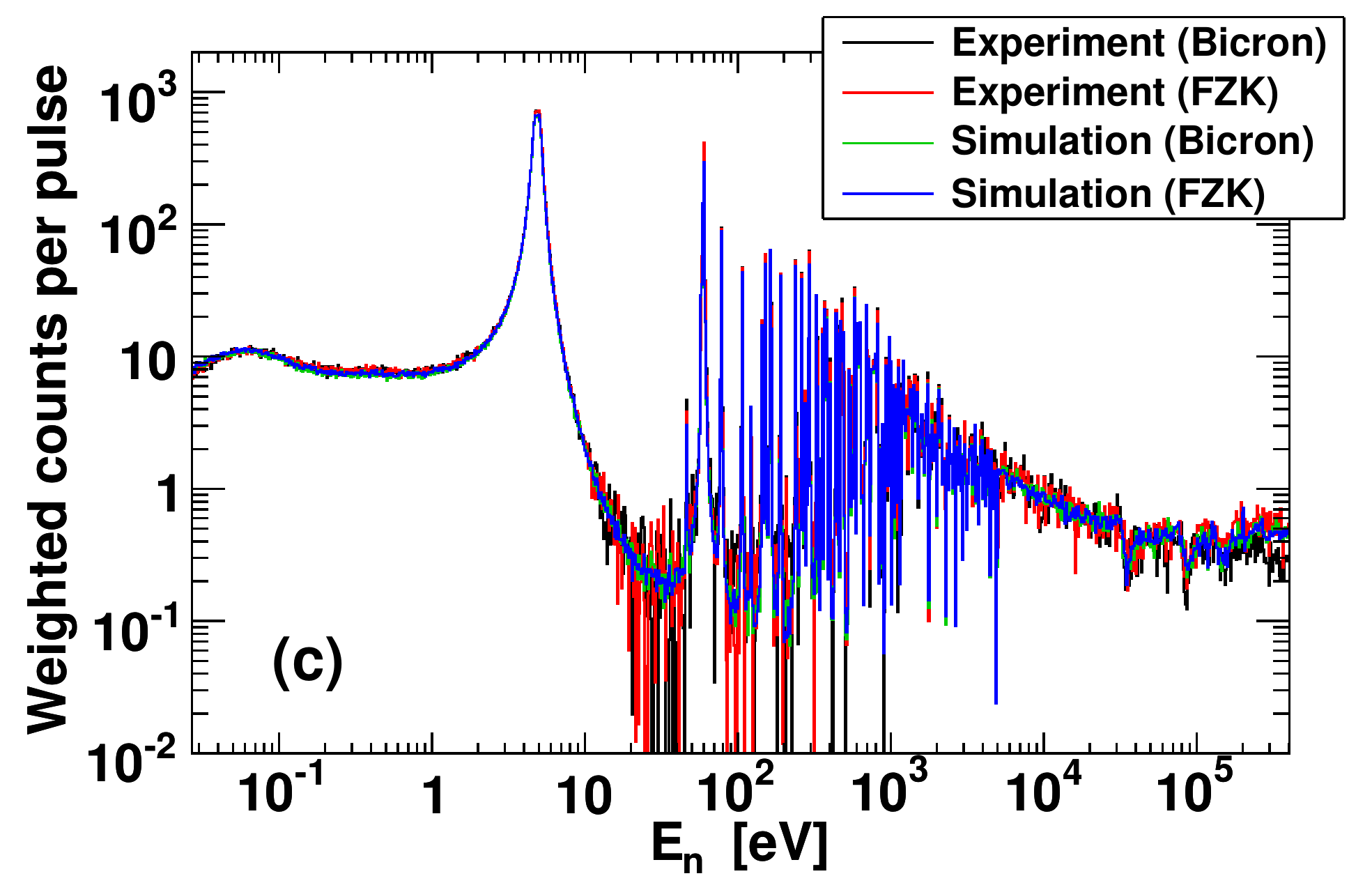}
\caption{(Color online) Radiative neutron capture of $^{197}$Au: comparison between the simulated and experimental results for the two C$_6$D$_6$ detectors, denoted as Bicron and FZK. Top panel (a): detector response to the $\gamma$-rays from the cascade following neutron capture. Middle panel (b): spectra of the neutron capture counts as a function of the neutron energy reconstructed from the neutron time-of-flight. Bottom panel (c): same as before after applying the weighting function technique.}
\label{fig6}
\end{figure}

The lack of correlations, and correspondingly the difference in the $\gamma$-ray cascade between simulations and tabulated data, could affect the reconstructed reaction yield. This, however, is not the case, thanks to the features of the PHWT. This is demonstrated by the analysis of the $^{197}$Au(n,$\gamma$) reaction. A thin $^{197}$Au sample is regularly irradiated during the experimental campaigns at n\_TOF, in order to provide the absolute normalization of the capture yield by means of the saturated resonance technique \cite{bibO}. Furthermore, $^{197}$Au is especially well suited due to its radiative capture cross section by far superseding the elastic scattering one in the energy range of the very large resonance at 4.9 eV. Therefore, it may be assumed with high confidence that the spectra, both experimental and simulated, are dominated by the detection of the $\gamma$-rays from the cascades following neutron capture. The comparison between the experimental and simulated $\gamma$-ray spectra for the two different C$_6$D$_6$ detectors used at n\_TOF, the Bicron and the FZK one, is shown in the top panel (a) of Fig. \ref{fig6}. A clear difference can be observed between the  simulated and measured spectra of energy deposition in both detectors, obviously related to the difference in the $\gamma$-ray cascade discussed above. The middle panel (b) shows the number of capture events as a function of neutron energy reconstructed from the time-of-flight. It can be noted that the simulated spectra differ considerably from their experimental counterparts, as expected from an energy dependent detection efficiency typical of the detectors. However, after the weighting functions have been applied to both the experimental and simulated data, as shown in the bottom panel (c) of Fig. \ref{fig6}, all spectra perfectly coincide. This is the exact purpose of the weighting function, i.e. to ensure that the detection efficiency of the employed detectors is independent of the cascade path. Although the $\gamma$-ray spectrum in the simulation shows differences relative to the tabulated one, as expected from the lack of correlation, these differences are practically eliminated in the final results by the PHWT. For this reason, the neutron transport capabilities of GEANT4 can be considered highly reliable, accurately reproducing the experimental data,  provided that the weighting function is taken into account in the analysis of the simulations, in a manner consistent with the analysis of the measurement. All results shown in Section \ref{sec:chap5} have been obtained under these conditions. 

\begin{figure}[b!]
\includegraphics[angle=90,width=1.\linewidth,keepaspectratio]{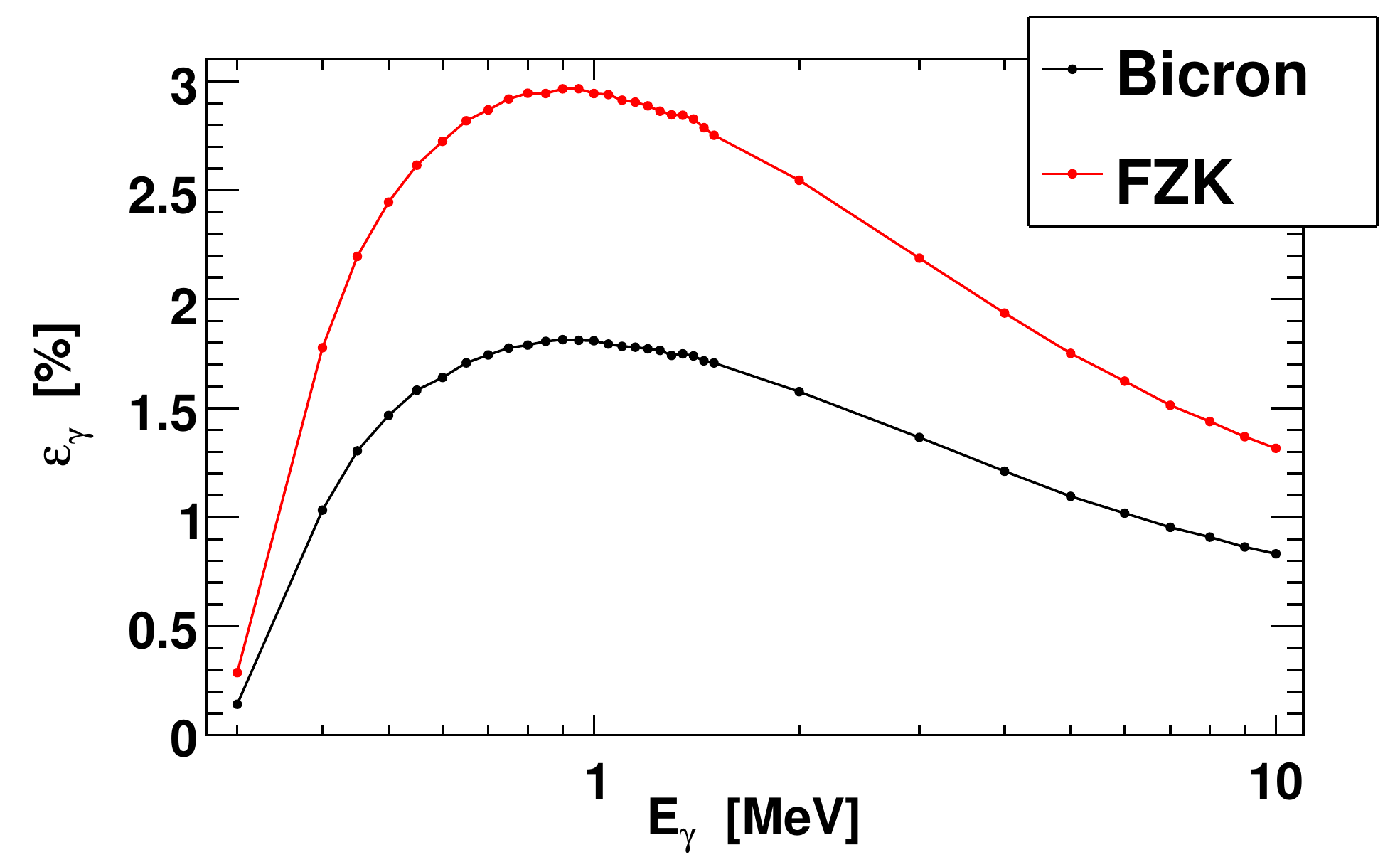}
\caption{(Color online) Total efficiency (intrinsic $\times$ geometric) of two C$_6$D$_6$ detectors for detecting $\gamma$-rays of given energy, isotropically emitted from a point source at the sample position. The maximal value for a given detector is used to define the neutron sensitivity of that detector.}
\label{fig17}
\end{figure}

\section{Neutron sensitivity}
\label{sec:chap4}

The conventional approach to the neutron sensitivity of an experimental setup consists in evaluating the ratio $\varepsilon_n/\varepsilon_\gamma^\mathrm{max}$ between the efficiency $\varepsilon_n$ for detecting a neutron, through the detection of secondary particles from the neutron-induced reactions, and the maximal $\gamma$-ray detection efficiency $\varepsilon_\gamma^\mathrm{max}$ of the setup. This procedure was adopted at n\_TOF for evaluating the neutron sensitivity of the optimized C$_6$D$_6$ detector, as reported in Ref. \cite{bibC}. The same definition is applied in this work to extract the neutron sensitivity of the whole setup. It should be noted, however, that the neutron energy range here considered extends from thermal to 1 MeV, i.e. five orders of magnitude more than in \cite{bibC}. Most importantly, while in the past only an isolated detector was studied, the present geometric configuration consists of the whole experimental setup, i.e. it includes the experimental hall with concrete walls as well as all other material present inside. In this respect, the simulations provide an overall neutron sensitivity of the setup.

In order to determine $\varepsilon_\gamma^\mathrm{max}$ for the two C$_6$D$_6$ detectors used at n\_TOF, a simulation was performed in which each detector was irradiated by monoenergetic $\gamma$-rays emitted isotropically from a point source at the sample position. A threshold of 200 keV was set on the energy deposited in a detector, replicating the experimental conditions at n\_TOF. The resulting detection efficiency $\varepsilon_\gamma$ is shown between 300 keV and 10 MeV in Fig. \ref{fig17} for both detectors. Contrary to Ref. \cite{bibC}, no solid angle corrections were applied, making $\varepsilon_\gamma$ the total detection efficiency, consisting of both the intrinsic and geometric component. The maximum efficiency is achieved for $\gamma$-rays of approximately 900 keV, being 1.8\% for Bicron, and approximately 3\% in case of FZK detector.

\begin{figure}[b!]
\includegraphics[angle=90,width=1.\linewidth,keepaspectratio]{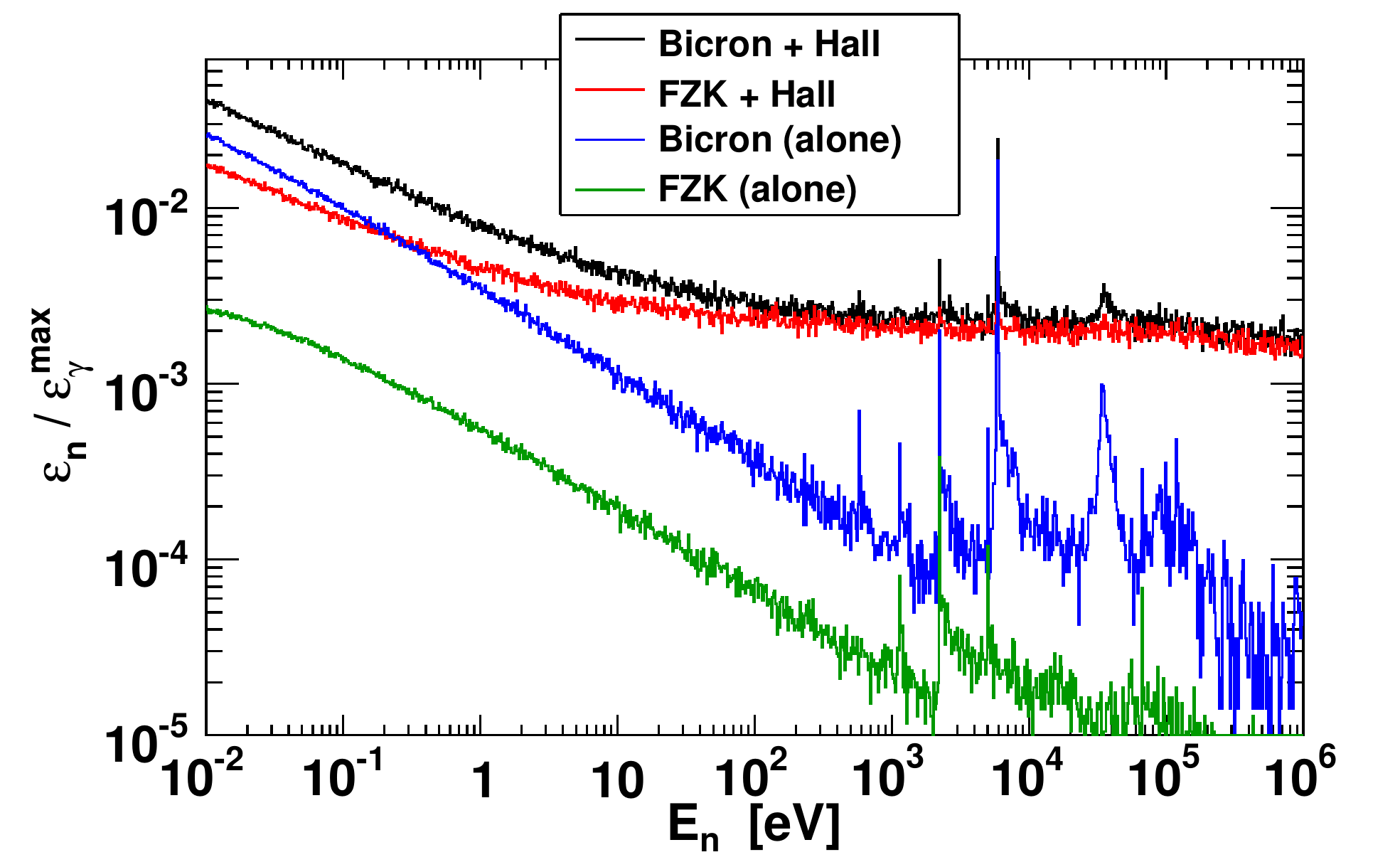}
\caption{(Color online) Neutron sensitivity of two C$_6$D$_6$ detectors in two configurations: detectors surrounded by the entire experimental hall (overall neutron sensitivity) and isolated detectors (intrinsic neutron sensitivity). The neutrons were emitted isotropically from a point source at the sample position, according to the $1/E_n$ energy distribution.}
\label{fig18}
\end{figure}

The neutron detection efficiency $\varepsilon_n$ was determined in the simulations in a similar way, i.e. irradiating the detectors with  neutrons emitted isotropically from a point source at the sample position. In this case as well, a threshold of 200 keV was set on the deposited energy. The primary neutron energy $E_n$ was sampled from an isolethargic distribution, that is uniform over the logarithm of energies (equivalently to a $1/E_n$ distribution). Two cases were considered in the simulations: in the first configuration, providing an intrinsic neutron sensitivity of the detectors, only an isolated detector was considered, consistently with the configuration adopted in Ref. \cite{bibC}. The second configuration, resulting in the overall neutron sensitivity, includes the whole setup, i.e. both detectors, the experimental hall and all other elements inside it. Since no solid angle corrections were considered in evaluating $\varepsilon_n$, as for $\varepsilon_\gamma^\mathrm{max}$, the neutron sensitivity determined from the ratio of the two efficiencies is implicitly corrected for the geometric contribution. The results for both detectors in the two different configurations are shown in Fig. \ref{fig18}. As expected, when considered separately, the FZK detector presents a reduced intrinsic neutron sensitivity relative to the Bicron one, mostly thanks to the  carbon fiber housing of the C$_6$D$_6$ liquid, contrary to the aluminum container used for the Bicron detector, which is also the cause of the prominent resonant structure above 1 keV. Another reason for the reduced intrinsic neutron sensitivity of the FZK detector is the higher value of $\varepsilon_\gamma^\mathrm{max}$, consequence of its larger active volume (1027 ml, as opposed to the 618 ml of the Bicron detector).

When considering the whole setup, the overall neutron sensitivity increases considerably for both detectors, exhibiting basically the same flat behavior above 10 eV, as shown in Fig. \ref{fig18}. The average value, over the range between 10 eV and 1~MeV is approximately $2\times10^{-3}$ for both detectors. It must be considered that, from the conventional definition, the neutron sensitivity represents the detector response to neutrons \emph{of} a given energy, and not \emph{at} the energy reconstructed from the neutron time-of-flight. As shown in the following Section, the only way to obtain a realistic estimate of the background related to the neutron sensitivity of the experimental setup and its time dependence, is to perform complete simulations including the specific sample being measured and the exact energy distribution of the neutron beam. Nevertheless, the neutron sensitivity defined in the conventional way provides important information on the expected magnitude of the background in a capture cross section measurement. More detailed information, however, in particular on the origin of this background, can be obtained by simulating the real case.

\begin{figure}[b!]
\includegraphics[angle=90,width=1.\linewidth,keepaspectratio]{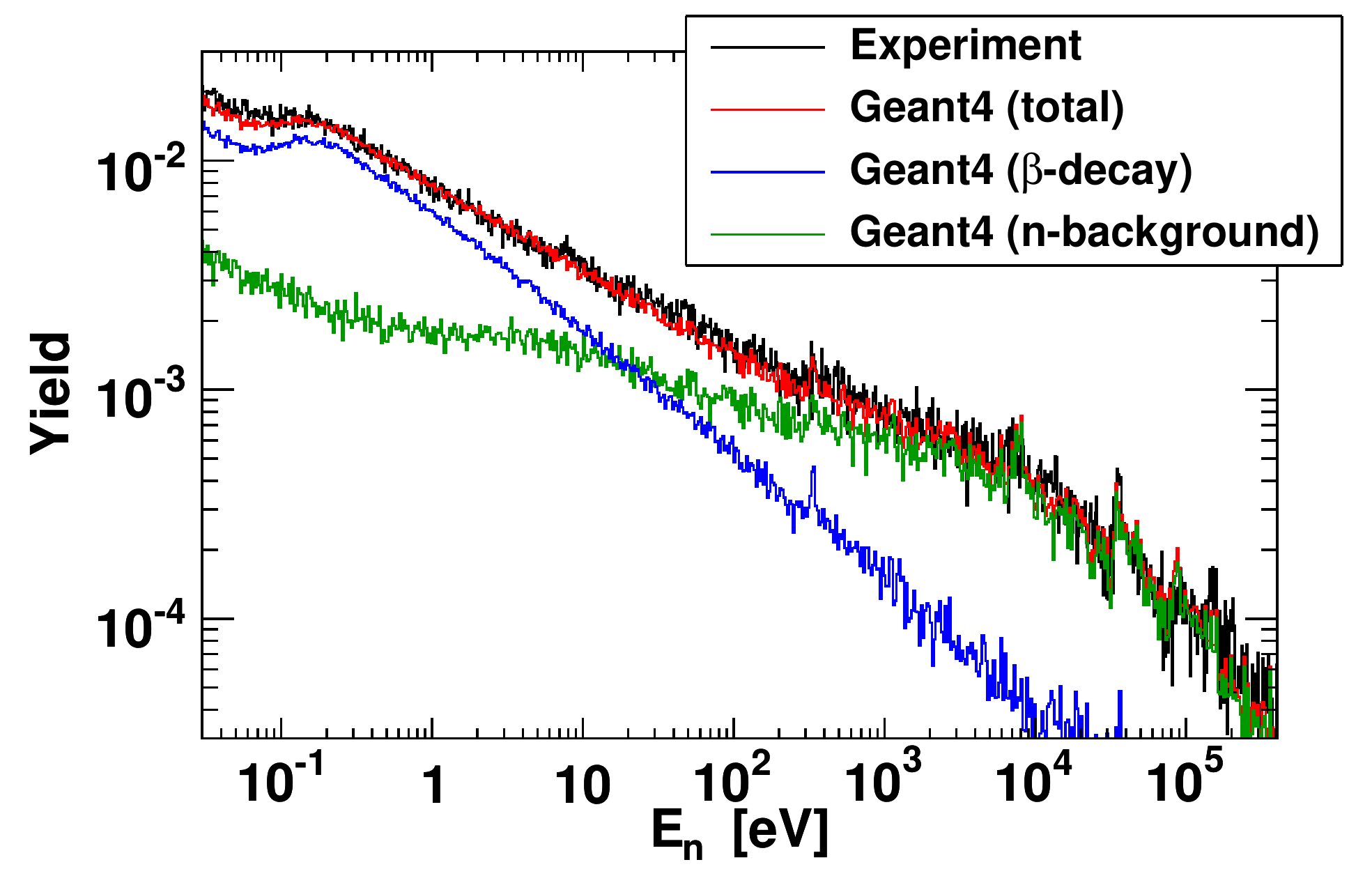}
\caption{(Color online) Comparison between the experimental and simulated yield for the $^\mathrm{nat}$C sample 1 cm in thickness and 2 cm in diameter. The contribution of the true neutron background -- caused by the scattering of the primary neutrons -- is shown separately from the detection of $\beta$-rays emitted by the radioactive residuals produced by neutron inelastic reactions in the sample.}
\label{fig7}
\end{figure}

\section{Neutron background}
\label{sec:chap5}

The GEANT4 simulations described above can be used to estimate the neutron background in a real measurement, related to neutrons elastically or inelastically scattered from the sample of a given isotope. Assuming that the scattering cross section for the isotope being measured is well known, the simulations are able to provide detailed information on the time structure of the background, or equivalently, its dependence on the neutron energy reconstructed from the time-of-flight. However, a validation of the GEANT4 results against experimental data is fundamental in order to ensure the reliability of the simulations and to estimate the associated uncertainty. To this end, ideally one would like to compare experimental data and simulations for a pure scatterer, i.e. an isotope with a capture cross section several orders of magnitude lower than for elastic scattering. A sample of $^\mathrm{nat}$C is one of the most appropriate choices in this respect, and for this reason it is routinely measured at n\_TOF as well as in other neutron facilities. In the following subsection, simulations performed for the neutron beam impinging on a thick $^\mathrm{nat}$C sample are compared with experimental data recently collected at n\_TOF, with the aim of validating the simulations. At the end of this Section, examples of neutron background affecting some other isotopes will be shown.

\begin{table}[t!]
\caption{List of reactions observed in the GEANT4 output data, producing $\beta$-radioactive residuals and contributing to the total $^\mathrm{nat}$C yield through the direct or indirect detection of $\beta$-rays. The isotopes from the upper part of the table are $\beta^-$-radioactive. The three carbon isotopes from the lower part of the table are $\beta^+$-radioactive. The $^{12}$B decay is by far the most important component in the measured yield.}
\label{tab1}
\begin{tabular}{cc>{\;}c>{\;}c}
\hline\hline
\textbf{Reaction}&\textbf{Isotope}&$\boldsymbol{\langle E_\beta\rangle}$&$\boldsymbol{t_{1/2}}$\\
\hline
$^{12}$C($n,p$)$^{12}$B&$^{12}$B&6.35 MeV&20.2 ms\\
$^{13}$C($n,p$)$^{13}$B&$^{13}$B&6.35 MeV&17.3 ms\\
$^{12}$C($n,p+\alpha$)$^{8}$Li&$^{8}$Li&6.20 MeV&840.3 ms\\
$^{12}$C($n,n+3p$)$^{9}$Li&$^{9}$Li&5.70 MeV&178.3 ms\\
$^{12}$C($n,p+d+\alpha$)$^{6}$He&$^{6}$He&1.57 MeV&806.7 ms\\
\hline
$^{12}$C($n,2n$)$^{11}$C&$^{11}$C&0.38 MeV&20.3 min\\
$^{12}$C($n,3n$)$^{10}$C&$^{10}$C&0.81 MeV&19.3 s\\
$^{12}$C($n,4n$)$^{9}$C&$^{9}$C&6.43 MeV&126.5 ms\\
\hline\hline
\end{tabular}
\end{table}

\subsection{Validation of the simulations with $^\mathrm{nat}\mathrm{C}$}

A high purity (99.95\%) $^\mathrm{nat}$C sample 1 cm in thickness and 2 cm in diameter was used to estimate the neutron background during a recent measurement of the capture cross section of $^{58}$Ni \cite{bibD}. The purity
of the $^\mathrm{nat}$C sample was checked by chemical analysis performed at Paul Scherrer Institute, and contamination by isotopes with high cross sections at thermal neutron energy was found to be below detection limit of few $\mu$g/g. Data collected on the $^\mathrm{nat}$C sample were taken as a benchmark for evaluating the accuracy of the GEANT4 simulation of the neutron background. In the simulations, a neutron beam with the same spectral and spatial characteristics of the n\_TOF beam impinged on a software replica of the $^\mathrm{nat}$C sample used in the measurement. The output data of the simulations were analyzed in exactly the same manner as the real data.

\begin{figure}[b!]
\includegraphics[angle=90,width=1.\linewidth,keepaspectratio]{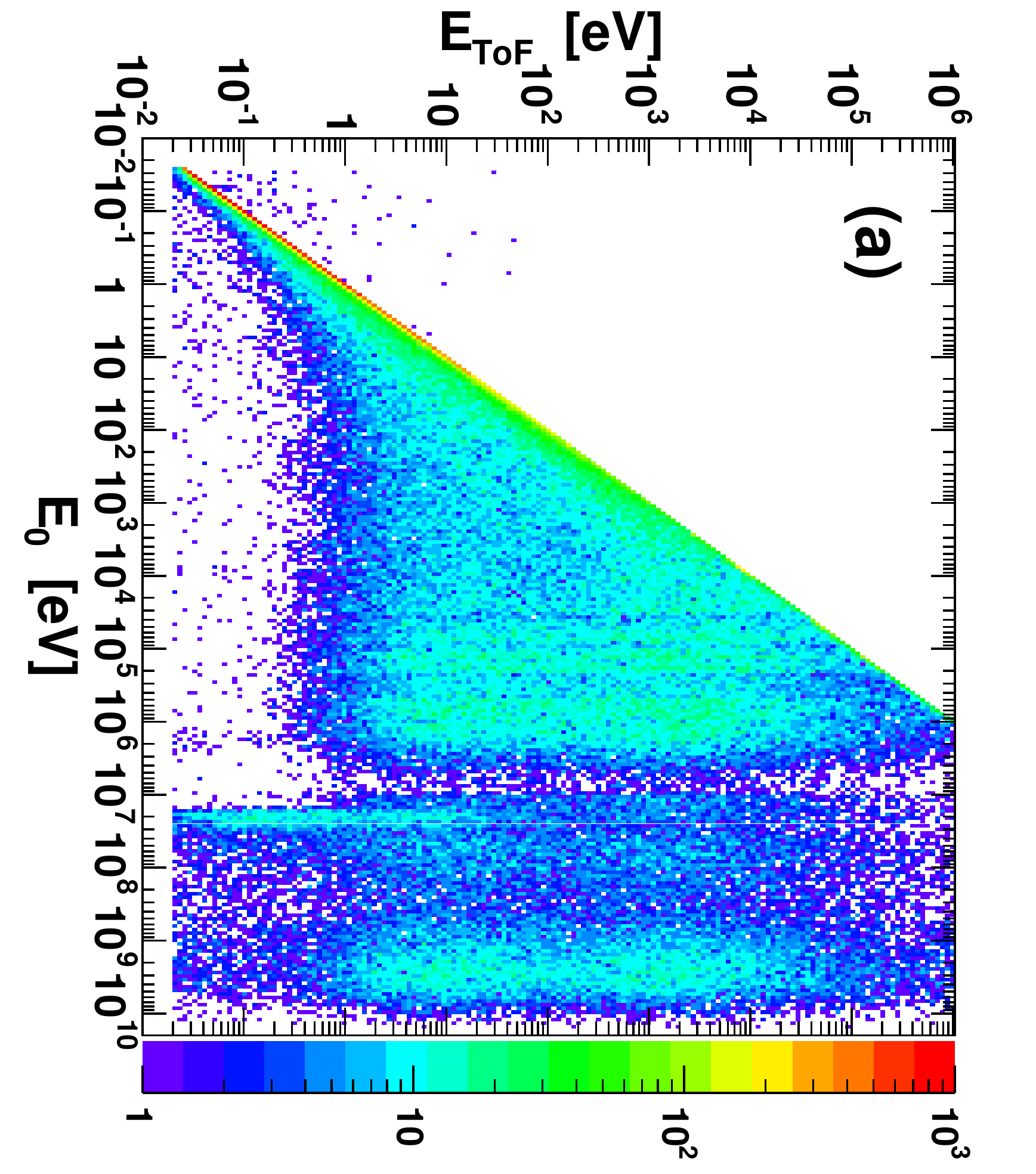}\\
\includegraphics[angle=90,width=1.\linewidth,keepaspectratio]{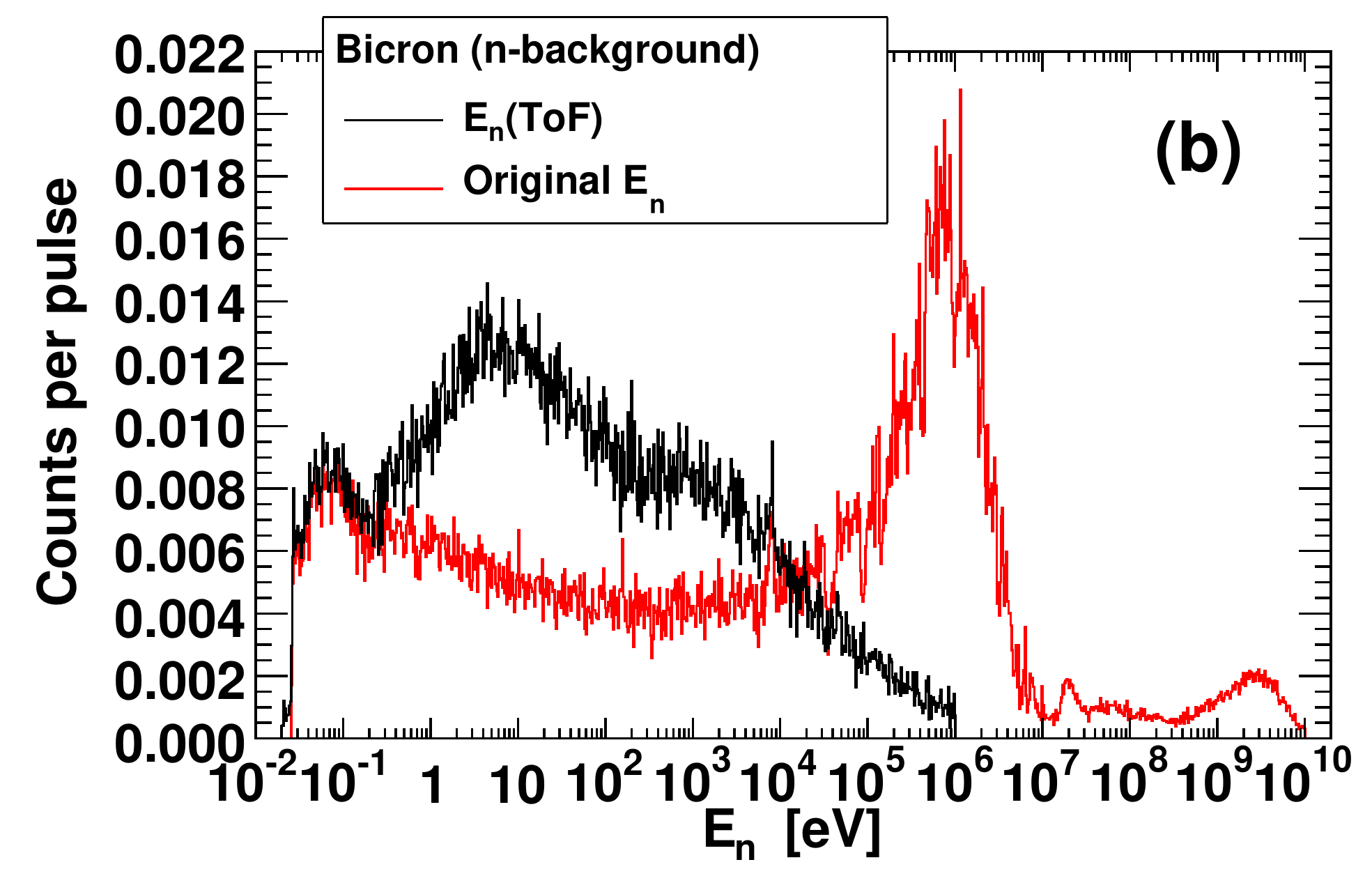}
\caption{(Color online) Time structure of the neutron background simulated for a $^\mathrm{nat}$C sample in the Bicron detector. Top panel (a): correlations between the primary neutron energies and those reconstructed from the time of flight. Bottom panel (b): projections of the top plot on both axes. Only the spectrum of reconstructed energies is experimentally accessible.}
\label{fig9}
\end{figure}

\begin{figure}[b!]
\includegraphics[angle=90,width=1.\linewidth,keepaspectratio]{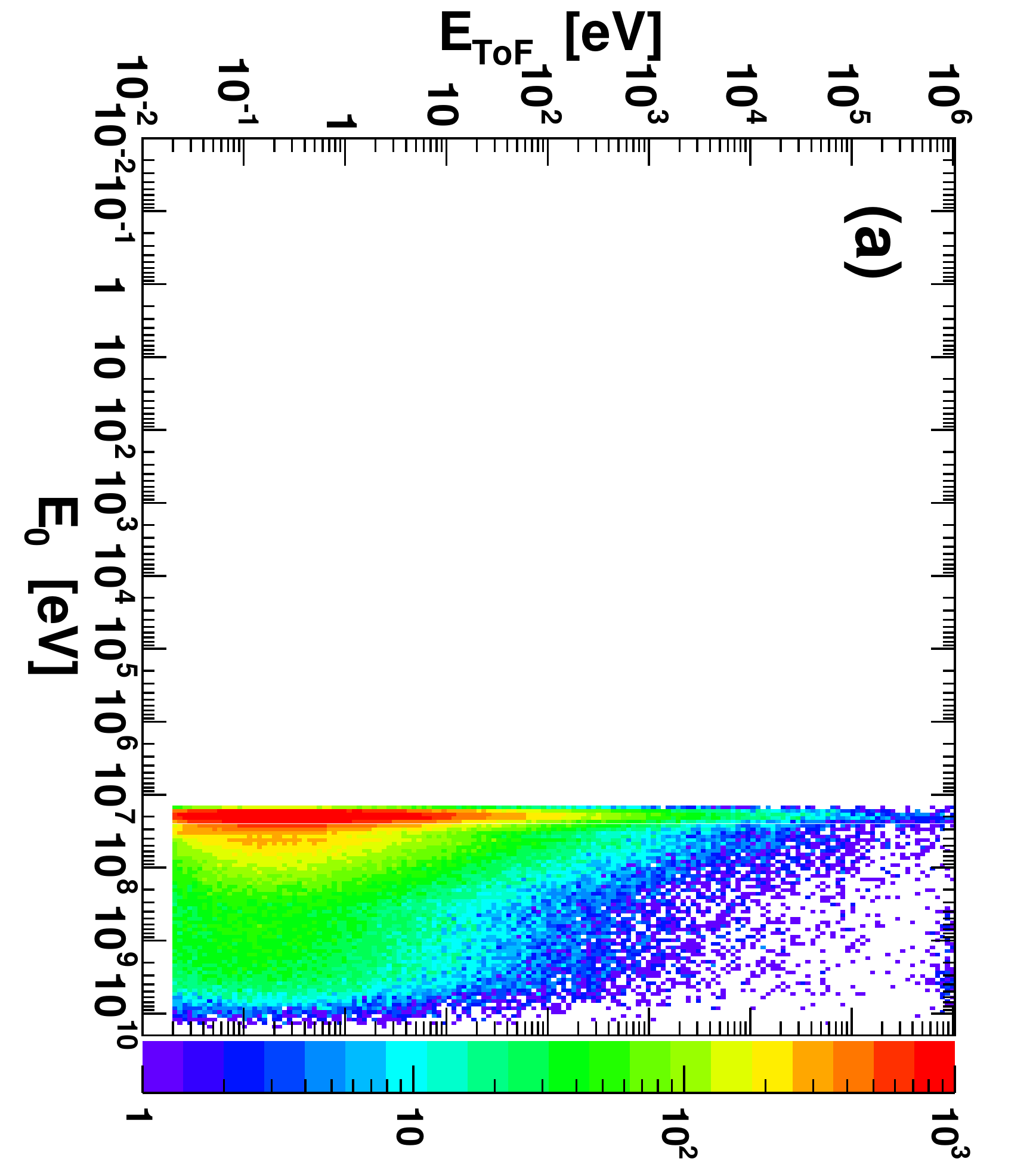}\\
\includegraphics[angle=90,width=1.\linewidth,keepaspectratio]{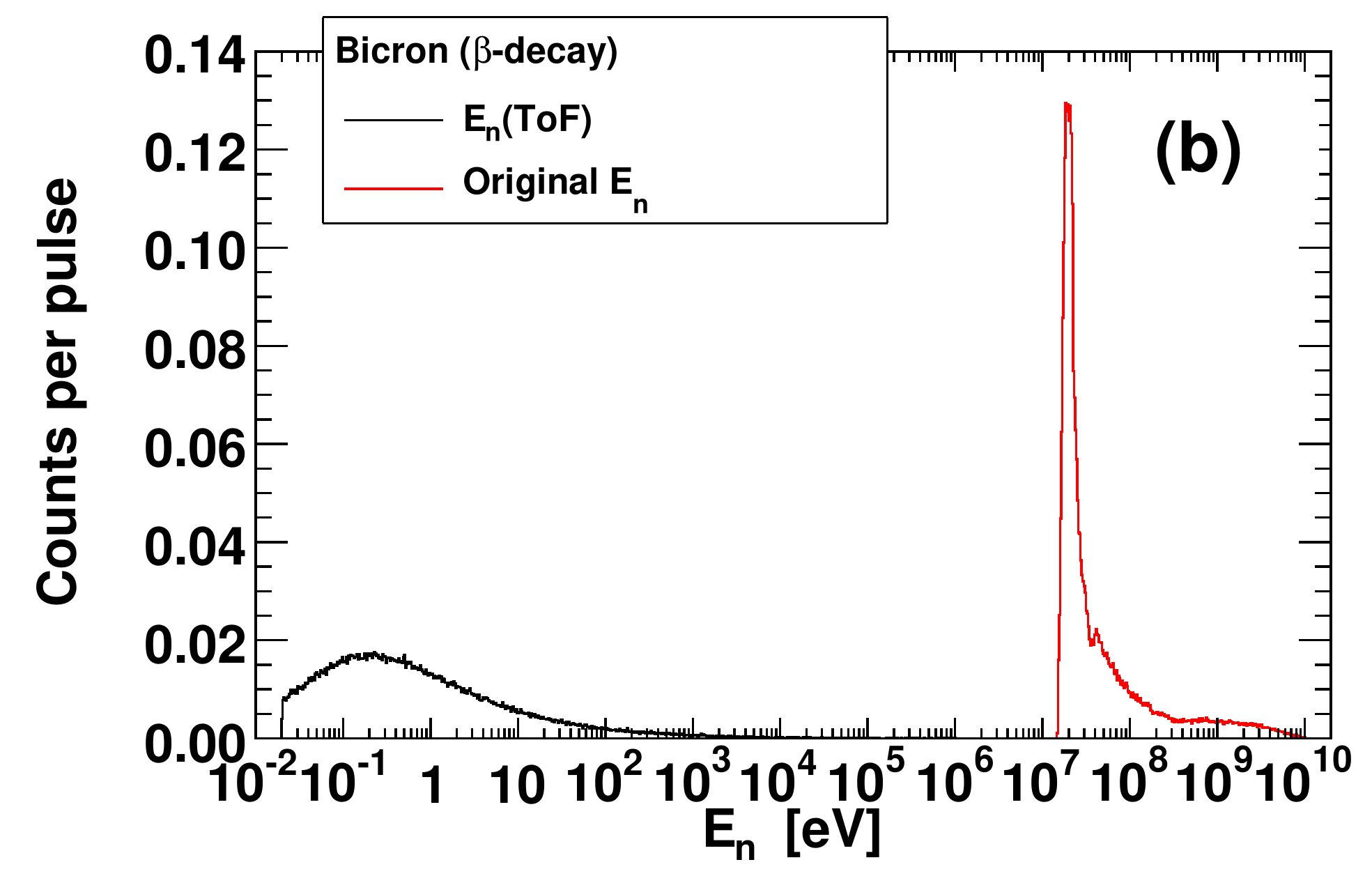}
\caption{(Color online) Time structure of the detected $\beta$-rays from the decay of radioactive isotopes produced by the neutron inelastic reactions in $^\mathrm{nat}$C sample. The plots are analogous to those from Fig. \ref{fig9}. The spectrum of primary neutron energies from bottom panel (b) bears the signature of the cross section for the dominant $^{12}$C($n,p$)$^{12}$B reaction.}
\label{fig10}
\end{figure}

Figure \ref{fig7} shows the comparison between the experimental yield (black histogram) and the simulated one (red histogram). Perfect agreement between measured and simulated yield can be observed  throughout the entire energy range from thermal up to 400 keV. However, a closer look at the simulations reveals an interesting effect. The simulation of the neutron background according to the usual definition, i.e. neutrons elastically scattered by the sample and captured anywhere in the experimental setup, is shown in Fig. \ref{fig7} by the green histogram. In this case, a reasonable agreement between the simulated neutron background and measured yield is only observed for reconstructed neutron energies above a few hundred eV, while a huge discrepancy, up to a factor of six, exists at low energy. The analysis of the simulations reveals that the excess yield in the measured histogram is in fact related to reactions induced by the high-energy part of the neutron beam on the $^\mathrm{nat}$C sample. In particular, neutrons of energy above a few MeV undergo ($n$,charged particle) reactions that produce short-lived $\beta$-emitters. Table \ref{tab1} lists examples of the observed reactions contributing to this effect, with the produced isotopes and their half-lives. By far the biggest contribution comes from the $^{12}$C($n,p$)$^{12}$B reaction. This reaction opens up at an incident neutron energy of approximately 15 MeV. The residual nucleus $^{12}$B is a $\beta^-$-emitter, with the mean electron energy of $\langle E_\beta\rangle$~=~6.35~MeV and a half-life of 20.2 ms. Being highly energetic, the electrons are able to reach the active volume of the C$_6$D$_6$ detectors, where they deposit a large fraction of their initial energy. In addition, due to the relatively short half-life of the decay, a significant portion of the exponential decay distribution falls within a 96 ms time-window used in the data acquisition (we remind here that the time-of-flight of a thermal neutron to the 187 m experimental area at n\_TOF is around 85 ms). In fact, the electron from the $\beta$-decay of $^{12}$B produces a signal at a "large" time-of-flight, corresponding to low reconstructed neutron energies. In other words, the $\beta$-decay being random in nature, the time-energy correlation for the primary neutron is completely lost, so that neutrons of energy above 15 MeV produce a signal in the thermal and epithermal region of the reconstructed neutron energy.

The contribution to the $^\mathrm{nat}$C yield from the ($n$,charged particle) reactions is shown in Fig. \ref{fig7} by the blue histogram. It should be mentioned that the exact magnitude of the effect depends on the cross section of the reactions, in particular the dominant $^{12}$C($n,p$)$^{12}$B reaction. Therefore, the combination of measured and simulated yield for $^\mathrm{nat}$C at n\_TOF can provide important indications on this rather uncertain $(n,p)$ cross section, integrated over a large neutron energy range (the uncertainty may be inferred from the discrepancies between the data available from various evaluated libraries \cite{bibP}).

Consequently, the contribution of high-energy ($n$,charged particle) reactions, which dominate the measured yield of $^\mathrm{nat}$C, questions the use of this sample for determining the true neutron background at low energies. Therefore, in capture studies at n\_TOF (and at similar facilities) neutron background from thermal to a few hundred eV can not be estimated from measurements with the $^\mathrm{nat}$C sample.

\begin{figure}[b!]
\includegraphics[angle=90,width=1.\linewidth,keepaspectratio]{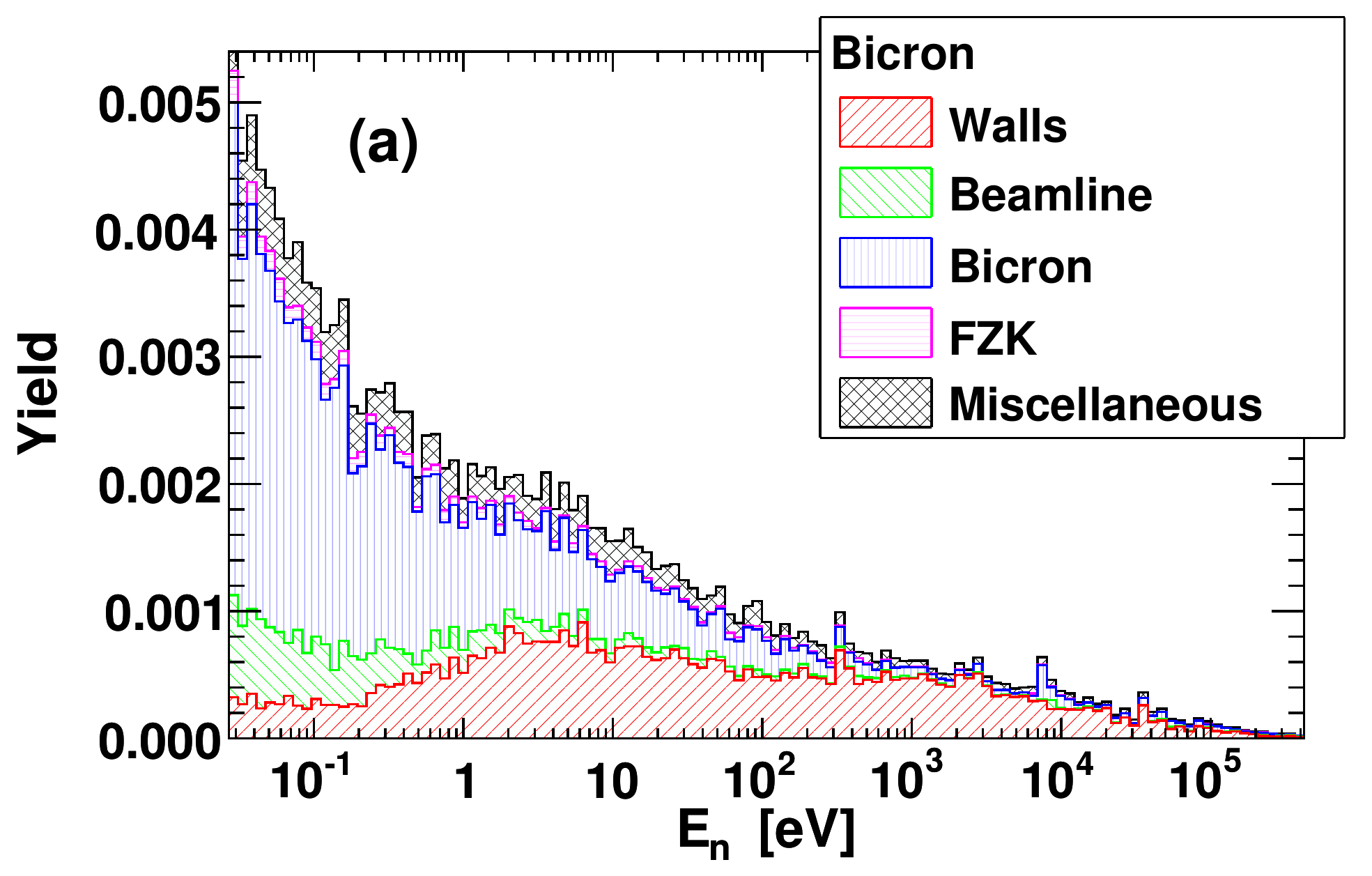}\\
\includegraphics[angle=90,width=1.\linewidth,keepaspectratio]{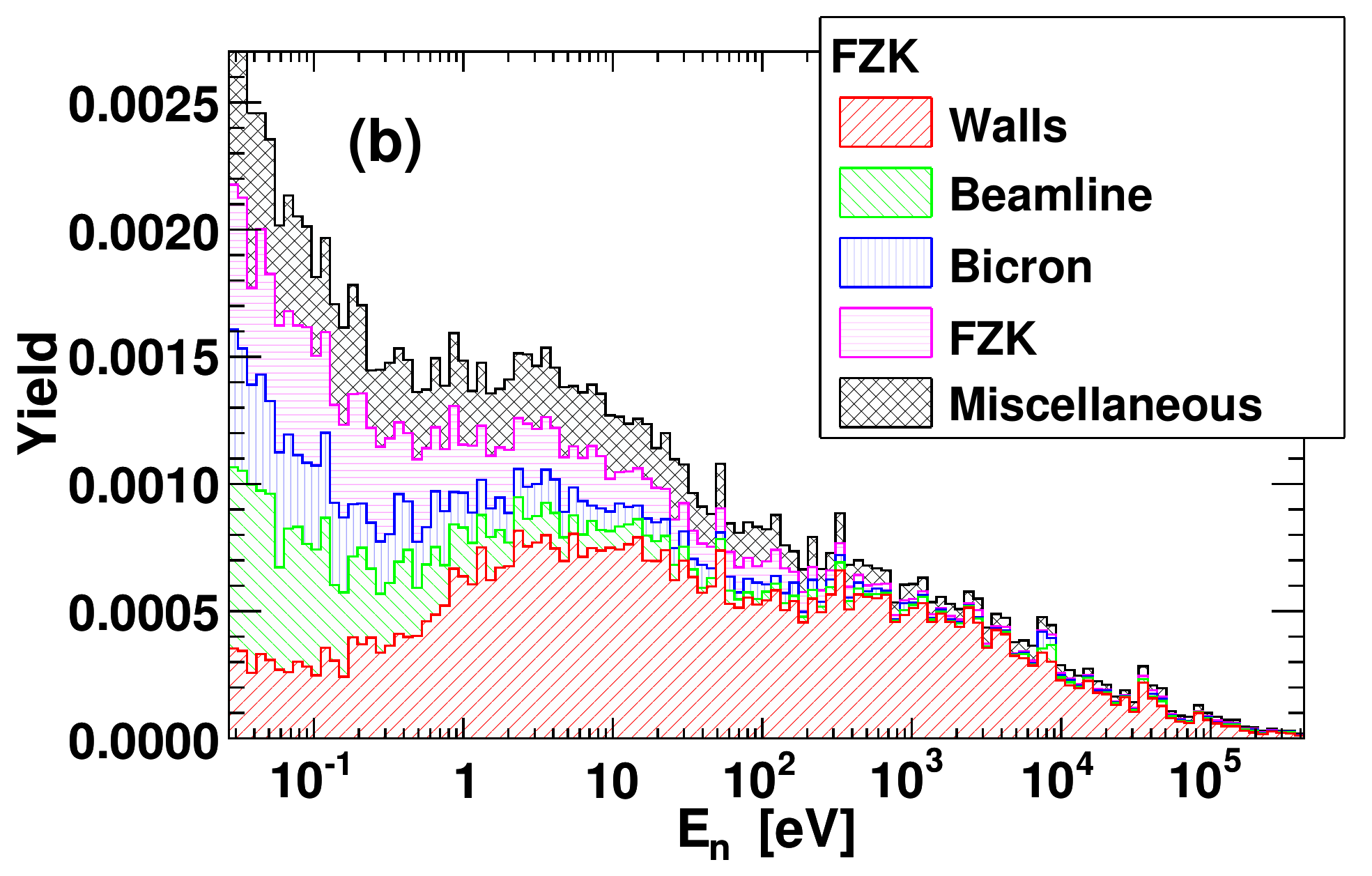}
\caption{(Color online) Stacked contributions to the neutron background yield of $^\mathrm{nat}$C for both C$_6$D$_6$ detectors -- Bicron (a) and FZK (b) -- coming from the different sets of experimental components. Note that absolute scales for two detectors differ by a factor of 2.}
\label{fig11}
\end{figure}

A further advantage of Monte Carlo simulations is that they allow one to study in detail the time structure of the neutron background, a quantity not easily accessible experimentally. A specific analysis of the simulations for the $^\mathrm{nat}$C sample has been performed to study this aspect and to localize the origin of the background. Such properties are expected to hold in general, i.e. for other samples as well, although the precise determination of the background, which depends on the scattering cross sections of the particular isotopes, can only be obtained by means of dedicated simulations covering the full energy range of the neutron flux.

An important issue that can be addressed with simulations regards the time structure of the neutron background. This problem is clearly summarized in Fig. \ref{fig9} (for simplicity, the figure refers only to the Bicron detector). The top panel (a) shows the correlations between the primary neutron energy $E_0$ and the energy $E_\mathrm{ToF}$ reconstructed from the total time-of-flight of the background event. The line $E_\mathrm{ToF}=E_0$ is the true prompt component, while all other counts show some degree of delay relative to the primary neutron. The bottom panel (b) of Fig. \ref{fig9} shows the projections of the results on the two axes. The data are shown only for the counts with the reconstructed energy $E_\mathrm{ToF}$ below 1 MeV. The significant loss of the correlations due to multiple neutron scattering inside the experimental hall is clearly evident from the shift in energy between the two spectra.

A final comment regards the presence in Fig. \ref{fig9} (a) (top panel) of some counts in the region above the 45 degree line (i.e. for a reconstructed energy higher than the original energy). These counts are related to the pulsed nature of the n\_TOF beam, reproduced in the analysis of the simulations. For some background events the time delay is so large that they show up in the next neutron pulse, where they are reconstructed with a much shorter time-of-flight. In particular, these counts in Fig. \ref{fig9} have been attributed to the decay of $^{28}$Al (with a half-life of 2.2 min), produced by activation of aluminum ($^{27}$Al) components surrounding the C$_6$D$_6$ detectors.

Figure \ref{fig10} shows the time structure of events related to the $\beta$-decay of $^{12}$B produced in the $^{12}$C($n,p$) reaction. The effect of the large delay associated with the $\beta$-decay is causing a huge shift from the primary neutron spectrum (15 MeV and higher) to the reconstructed energy in the thermal and epithermal region.

Some considerations can be made on the origin of the background on the basis of Fig. \ref{fig11}, which shows the contribution of the various components of the experimental setup to the total neutron background. The biggest contributions are related to the walls of the experimental hall, the beamline tubes, the various parts of both detectors, while all remaining components account for a small fraction of the background. In particular, the walls of the experimental area are the most significant source of neutron background above 1 keV, since the walls are very effective in moderating higher energy neutrons. Secondly, beamline tubes are the prominent source of background at lower energies due to the aluminum composition. Some part of the beamline is also very close to the detectors, which enhances their contribution. The use of carbon fiber instead of aluminum would, therefore, be preferable. Finally, the figure clearly shows the large contribution coming from the aluminum housing of the Bicron detector, in particular on the neutron background affecting the Bicron detector itself, although some contribution is observed also for the FZK facing it in the simulated experimental setup.

\subsection{Examples of neutron background: $^\mathrm{197}\mathrm{Au}$ and $^\mathrm{58}\mathrm{Ni}$}

\begin{figure}[t!]
\includegraphics[angle=90,width=1.\linewidth,keepaspectratio]{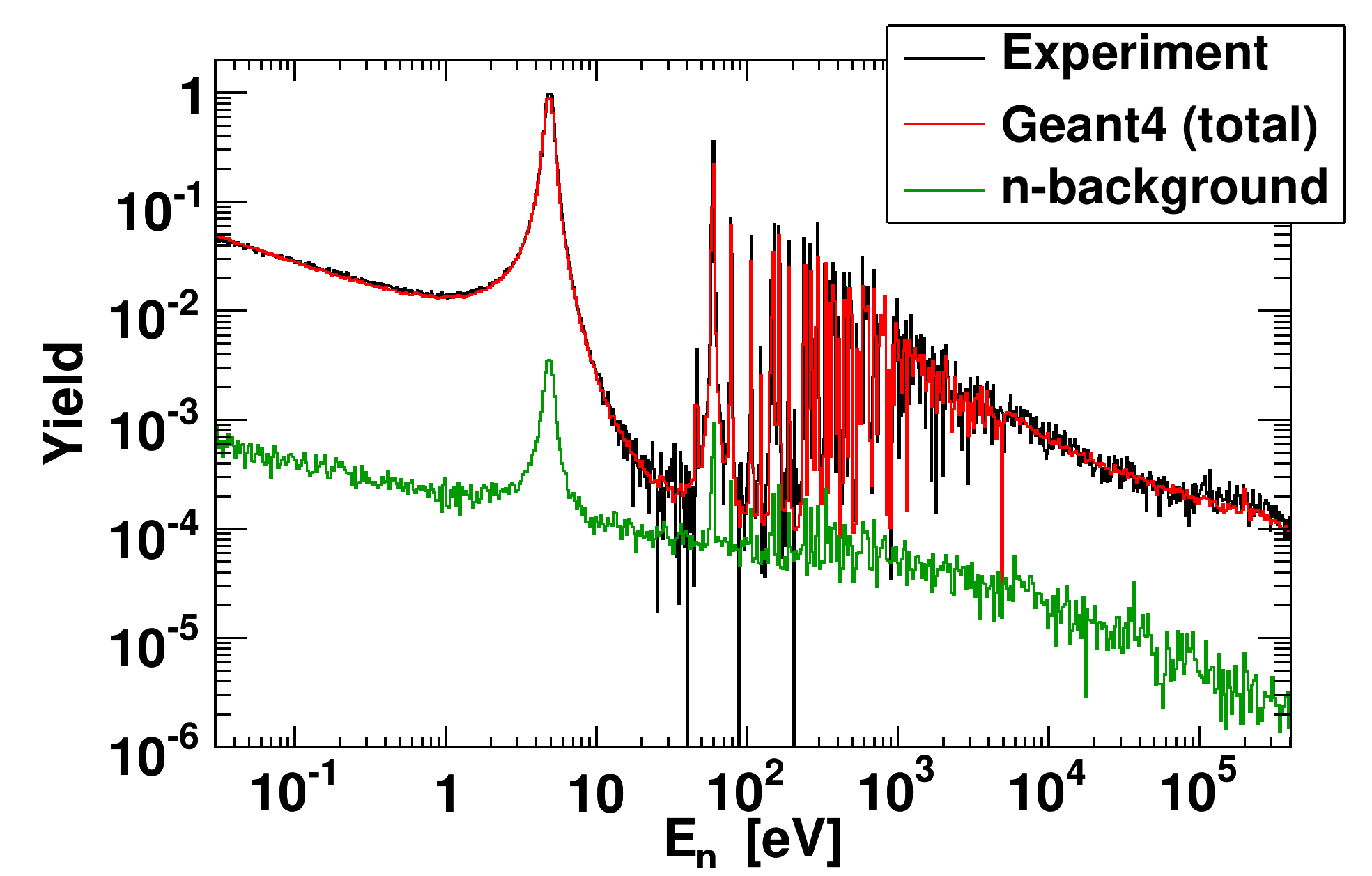}
\caption{(Color online) Comparison between the experimental and simulated neutron capture yield of $^{197}$Au. The neutron background is also shown.}
\label{fig14}
\end{figure}

The correct estimate of the neutron background is important for an accurate determination of the capture cross section, in particular for isotopes characterized by a large scattering-to-capture cross section ratio. To show the influence on the neutron background in capture cross section measurements, it is instructive to consider two examples. $^{197}$Au is routinely measured in capture experiments for the purpose of absolute normalization of the cross section. This isotope is characterized by a large capture cross section, in particular at low energy, so the neutron background should not play a big role. This is in fact the case as shown in Fig. \ref{fig14}. The measured total yield (in black) is here compared with the simulated one (in red) (for these examples, capture reactions are also included in GEANT4 simulations). The simulated neutron background is also shown in the figure. As expected, the neutron background is small, of the order of a few percent, except in the resonance valleys. In particular, in the 4.9 eV resonance used for normalization with the saturated resonance method, the neutron background is at the level of less than a percent, and can safely be neglected in the analysis.

\begin{figure}[t!]
\includegraphics[angle=90,width=1.\linewidth,keepaspectratio]{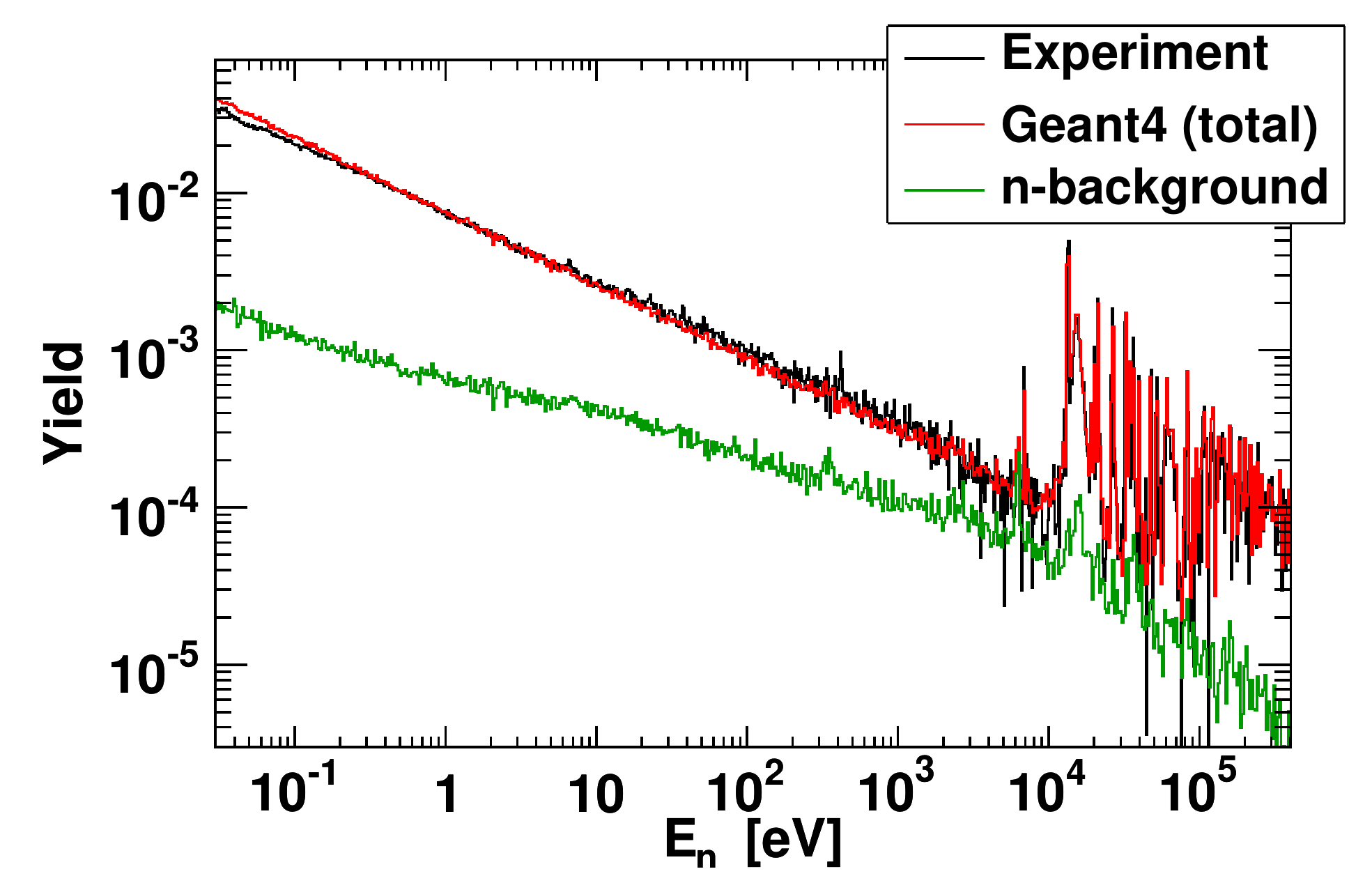}
\caption{(Color online) Comparison between the experimental and simulated neutron capture yield of $^{58}$Ni. The neutron background is also shown.}
\label{fig16}
\end{figure}

As illustrated in Fig. \ref{fig16}, the situation for $^{58}$Ni, which was recently measured at n\_TOF \cite{bibD}, appears to be completely different. Below approximately 1 keV the total experimental yield is dominated by the simulated neutron background and the effect in the keV region is also much larger than for $^\mathrm{197}\mathrm{Au}$. It is, therefore, mandatory in this case to take it into account, in order to derive the true capture yield. For the first time, the present simulations allowed us to determine the neutron background in the whole energy range and with the correct time-structure, thus leading to more accurate capture cross sections compared to previous measurements, where the neutron background was either neglected or roughly estimated.

\section{Conclusions}
\label{sec:chap6}
 
C$_6$D$_6$ detectors are routinely employed in measurements of radiative neutron capture cross sections at the neutron time-of-flight facility n\_TOF at CERN, as well as at other neutron facilities around the world, because of their very low intrinsic neutron sensitivity. This characteristic is fundamental in minimizing the background related to neutrons scattered by the sample under investigation. However, the knowledge of the neutron sensitivity of the detectors alone is not sufficient for a precise estimate of the neutron background, since part of it could be produced by the interaction of scattered neutrons with the whole experimental set-up, including the experimental hall. A realistic estimate of the overall neutron background in a wide energy range can be obtained by measuring a $^\mathrm{nat}$C sample, which acts as a pure neutron scatterer. However, such a measurement does not provide information on the time structure of the background and could be affected by other effects at low neutron energies. 

In this paper, the neutron sensitivity of the whole capture setup used at n\_TOF has been studied by means of Monte Carlo simulations performed with the GEANT4 tool. A complete and detailed software replica of experimental hall and apparatus has been implemented in the simulations, and the results have been analyzed with the same procedure applied to the experimental data. 
All relevant physical processes were taken into account in the simulations, for neutrons generated in the full energy range of the n\_TOF facility, i.e. from thermal up to 10 GeV. The lack of correlations between $\gamma$-rays in a cascade, typical of most Monte Carlo simulations, has been proven to bear no effect on the results, provided that the weighting function technique is correctly applied to the simulations.

The simulations have been validated by comparing the background simulated for the neutron beam impinging on a 
$^\mathrm{nat}$C sample against experimental data. Remarkable agreement is found for neutron energies in the range between 1 keV and 1 MeV, demonstrating that highly reliable results for the neutron background can be obtained by means of complete GEANT4 simulations. A close investigation of the Monte Carlo results below 1 keV neutron energy has led to the discovery of an additional component related to the $\beta$-decay of $^{12}$B produced in the $^{12}$C($n,p$) reaction. Although this reaction occurs for neutron energies above 15 MeV, it constitutes by far the dominating contribution to the measured yield of $^\mathrm{nat}$C at low energy, as a consequence of the 20.2 ms decay time of $^{12}$B, and of the high energy of the electrons. It is important to remark that this component could be present in the $^\mathrm{nat}$C measurement at any other time-of-flight facility in which the energy range of the neutron beam extends beyond 15 MeV.

The wealth of data extracted from the simulations has provided information, experimentally unaccessible, on the origin and time-dependence of the background. In particular, the analysis of the GEANT4 simulations on the neutron sensitivity has led to the following conclusions:
\begin{itemize}
\item[$\bullet$] The neutron sensitivity of the whole capture setup is sensibly larger than for the detectors alone.
\item[$\bullet$] At neutron energies below a few keV the main contribution to the overall neutron background comes from the detectors and the beam-line components, whereas the contribution from the walls of the experimental area is dominating at higher energies.
\item[$\bullet$] An important loss of correlation in time is observed between incident neutron energy and detected background, in particular above a few eV. The time structure of the background needs to be taken into account in particular when estimating the neutron background below resonances, in particular in the keV region. Monte Carlo simulations are at present the only means for estimating the time dependence of the background.
\end{itemize}

Finally, examples of the neutron background for two cases, $^{197}$Au and  $^{58}$Ni, have been presented, demonstrating the importance of the simulations in correctly estimating and subtracting the neutron background in measurements on isotopes exhibiting a large scattering-to-capture cross section ratio.\\

\textbf{ Acknowledgements}\\

We are indebted to the members of Laboratory for Advanced Computing at Faculty of Science, University of Zagreb, for granting us access to the computer cluster we have used for running the simulations. The authors are indebted to the national and international
funding agencies that have supported the n\_TOF Collaboration. The research leading to these results has received funding from the European Atomic Energy Community's (Euratom) Seventh Framework Programme FP7/2007-2011 under the Project "ANDES" (Grant Agreement n 249671) and CHANDA (GA n. 605203).

\end{document}